\documentclass[aps,prb,twocolumn,superscriptaddress,floatfix,nofootinbib]{revtex4-2}
\usepackage{epsfig,amsmath,amssymb,color,comment,physics}
\usepackage[makeroom]{cancel}
\usepackage[caption=false]{subfig}
\usepackage{mathrsfs}
\usepackage[countmax]{subfloat}
\usepackage[normalem]{ulem}
\usepackage[english]{babel}
\usepackage{dsfont}
\usepackage{float}
\usepackage[bookmarks=true,colorlinks,linkcolor=black,urlcolor=NavyBlue,citecolor=RoyalBlue]{hyperref}
\usepackage[dvipsnames]{xcolor}
\usepackage{soul} 
\usepackage{color, xcolor} 

\usepackage{color}

\newcommand{\beginsupplement}{%
    \setcounter{table}{0}
    \renewcommand{\thetable}{S\arabic{table}}%
    \setcounter{figure}{0}
    \renewcommand{\thefigure}{S\arabic{figure}}%
    \setcounter{equation}{0}
    \renewcommand{\theequation}{S\arabic{equation}}%
    \setcounter{section}{0}
    \renewcommand{\thesection}{S\arabic{section}}%
   }

\graphicspath{{./Figures/}} 

\begin{document}

\title{Few-Body Quantum Chaos, Localization, and Multi-Photon Entanglement in Optical Synthetic Frequency Dimension}

\author{Junlin Wang} 
\affiliation{School of Physics, Zhejiang Key Laboratory of Micro-nano Quantum Chips and Quantum Control, Zhejiang University, Hangzhou $310027$, China}

\author{Luojia Wang} 
\affiliation{State Key Laboratory of Advanced Optical Communication Systems and Networks, School of Physics and Astronomy, Shanghai Jiao Tong University, Shanghai $200240$, China}

\author{Jinlou Ma} 
\affiliation{School of Physics, Zhejiang Key Laboratory of Micro-nano Quantum Chips and Quantum Control, Zhejiang University, Hangzhou $310027$, China}

\author{Ang Yang} 
\affiliation{School of Physics, Zhejiang Key Laboratory of Micro-nano Quantum Chips and Quantum Control, Zhejiang University, Hangzhou $310027$, China}

\author{Luqi Yuan} \email{yuanluqi@sjtu.edu.cn}
\affiliation{State Key Laboratory of Advanced Optical Communication Systems and Networks, School of Physics and Astronomy, Shanghai Jiao Tong University, Shanghai $200240$, China}

\author{Lei Ying} \email{leiying@zju.edu.cn}
\affiliation{School of Physics, Zhejiang Key Laboratory of Micro-nano Quantum Chips and Quantum Control, Zhejiang University, Hangzhou $310027$, China}

\date{\today}

\begin{abstract}
Generation and control of entanglement are fundamental tasks in quantum information processing. In this paper, we propose a novel approach to generate controllable frequency-entangled photons by using the concept of synthetic frequency dimension in an optical system. Such a system consists of a ring resonator made by a tailored third-order nonlinear media to induce photon-photon interactions and a periodic modulator to manipulate coupling between different frequency modes. We show this system provides a unique platform for the exploration of distinct few- or many-body quantum phases including chaos, localization, and integrability in a highly integrable photonics platform. In particular, we develop the potential experimental method to calculate the spectral form factor, which characterizes the degree of chaos in the system and differentiates between these phases based on observable measurements. Interestingly, the transition signatures of each phase can lead to an efficient generation of frequency-entangled multi photons. This work is the first to explore rich and controllable quantum phases beyond single particle in a synthetic dimension.
\end{abstract}
\maketitle

{\bf Introduction---}Realization of entanglement by photons is one of important sources in the realm of quantum technology, spanning applications from quantum computing and teleportation to secure communication, quantum swapping, and beyond~\cite{jozsa2003role,Bouwmeester1997,gao2005deterministic,su2016quantum}. Among them, generation of quantum entangled states across various dimensions holds immense potential for specific fields. 
For example, entangled states with long distance in spatial dimensions find practical utility in secure communication~\cite{shannon2020use,long2007quantum,qi2019implementation,portmann2022security,paraiso2021photonic}. 
Also, entangled states in other degrees of freedom or dimensions are attracting significant interest due to their potential applications~\cite{Mair2001,Defienne2021,Cheng2023}.

In recent years, the concept of synthetic dimensions in photonics has emerged as an artificial space that harnesses the degrees of freedom of light, such as frequency~\cite{Yuan:16,PhysRevA.93.043827}, spatial mode~\cite{Lustig2019}, or orbital angular momentum~\cite{Luo2015} of light to provide a novel platform for emulating numerous models in various systems~\cite{Javid2023,Xiao2022,PhysRevLett.128.223602}, and hence stands out as a promising tool for artificially generating and manipulating the systematic parameter in photon entanglement generations. On the other hand, the introduction of nonlinear optical materials to optical structures enables photons to interact with each other through the media. This facilitates the study of photon-photon interaction problems in synthetic frequency dimensional systems~\cite{yuan_creating_2020,PhysRevA.102.023518,Englebert2023}, a departure from the typical focus on the single-particle problem in related synthetic dimension studies~\cite{Yuan:18,Yuan2021,Heckelmann2023,Senanian2023,PhysRevLett.130.143801,PhysRevLett.130.083601,Li2023,Cheng2023}. Consequently, a question arises: can the build of artificial dimension be effectively applied in the many-body limit, at least in the few-body regime, to generate and control frequency-entangled states of photons?

\begin{figure}
\centering
\includegraphics[width=\linewidth]{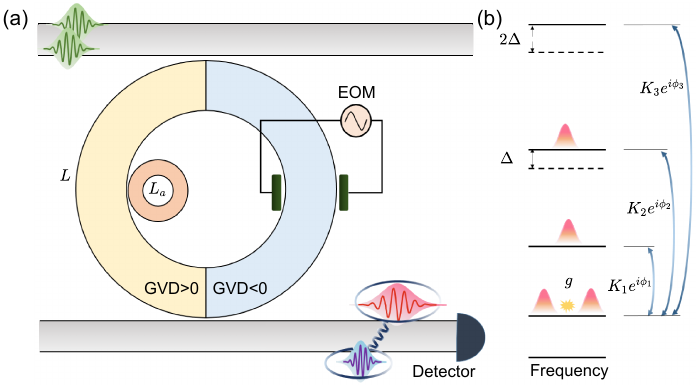}
\caption{
(a) Schematic diagram of a system consisting of a ring resonator and two waveguides. The ring resonator is comprising two types of nonlinear media (yellow and blue) with the opposite GVD and the same $\chi^{(3)}$ and it is dynamically modulated. An auxiliary resonator effectively creates open boundary conditions in the synthetic frequency dimension~\cite{dutt_creating_2022}. The system is initialized with two single photons (green pulses) and then such a system generates two entangled photons (red and blue pulses) at the output. (b) Photons interaction model in synthetic frequency lattice created by periodic boundary conditions and dynamic modulation.  
  }\label{fig:image1}
\end{figure}

Here, building on the work about photon-photon interactions in Ref.~\cite{yuan_creating_2020}, we take an essential step towards investigating the supporting quantum phases, phase transitions, and then transition signatures for generating novel entangled states with the synthetic frequency dimension.
 We strictly derive an extended quantum many-body model in the synthetic frequency dimension by strictly quantizing the electromagnetic field within a ring resonator.
This system can simulate a frequency-level ``chain'' with photon-photon interactions and long-range couplings.  By adjusting parameters, we achieve controllable transitions between distinct quantum phases, including quantum chaos~\cite{Nation_2018,gomez2011many,PhysRevX.8.021062,stockmann2000quantum}, Stark many-body localization (SMBL)~\cite{pal2010many,nandkishore2015many,smith2016many,abanin2019colloquium,alet2018many} , and integrable phases. Those phases can be predicted by the spectral form factor (SFF) from observable measurements~\cite{PhysRevE.106.024208,joshi_probing_2022,PhysRevX.8.021062}.
Remarkably, we exploit the unique properties of quantum chaos and SMBL to generate frequency-entangled photon states beyond the single-particle limit~\cite{PhysRevLett.122.083903}.

{\bf Models---}As shown in Fig.~\ref{fig:image1}(a), we consider a ring resonator coupled to upper and lower waveguides, which manage the input and output of photons, respectively.
The ring itself comprises two nonlinear media: a left- and a right-hand media, each possessing non-zero third-order nonlinear coefficients and exhibiting opposite group velocity dispersion (GVD)~\cite{yuan_creating_2020}. 
Its perimeter $L$ is much larger than the photon wavelength $\lambda$ and thickness of the ring $d$. A phase modulator is coupled to the ring resonator for dynamically tunning the phase. For an ideal ring without any leakage to environment, the Hamiltonian of the quantum electromagnetic field in this nonlinear ring is given by
\begin{equation}\label{eq1}
\begin{split}
\hat{H}_0=\frac{1}{2}&\int dv \left(\epsilon_{m}  \hat{E}^2 + \frac{1}{\mu_{m}}\hat{B}^2\right) \\
+ &\int dv \sum_{nmpq} \chi^{(3)} \hat{E}_n \hat{E}_m \hat{E}_p \hat{E}_q,
\end{split}
\end{equation}
where $\chi^{(3)}$ is the third-order nonlinear coefficient and $\epsilon_{m} = \epsilon_{0}\epsilon_{r}$ and $\mu_{m} = \mu_{0}\mu_{r}$ are the permittivity and permeability of the ring resonator media, respectively. Here, $\epsilon_0$ and $\mu_0$ are the vacuum permittivity and permeability, respectively. 
The electromagnetic field operators $\hat{E} = \hat{F}^{(1)} $ and $\hat{B} = \hat{F}^{(2)}$ in such a ring structure are written as:
\begin{equation}\label{eq2}
\begin{split}
&\hat{F}^{(1,2)}(z,t)
=    \sum_n  F_n^{(1,2)}
=  \sum_{n} C^{(1,2)}_n  \\
&\times\Bigg[ \hat{a}_n  {\rm sin} \left(k_n + \frac{\phi (t)}{L}\right) l  
 + b^{(1,2)}\hat{a}_n^\dagger   \mathrm{sin} \left(k_n + \frac{\phi (t)}{L}\right) l\Bigg],
\end{split}
\end{equation}
where $C_n^{(1)} = i\sqrt{ {h \omega_n}/({2 V \epsilon_m}) }$, $C^{(2)}_n = \sqrt{ {h \omega_n}/({2  \epsilon_m c^2 V}) }$, $b^{(1,2)}=-1,+1$, and $k_n= 2 n \pi / L$ with integer $n$ being the mode index of the ring resonator.  $c$ is the speed of light and $V$ is the effective model volume. $\hat{a}_n$ and $\hat{a}_n^\dagger$ represent the creation and annihilation operators of the $n$th mode, respectively.
The phase modulation is written as~\cite{Optical_Electronics_in_Modern}
$\phi(t) = \sum_{\mu=1}^{M} \alpha_{\mu}\left[ e^{i\left( \Omega_{\mu} t+\phi_\mu \right) } + e^{-i\left( \Omega_{\mu} t+\phi_\mu \right) }\right] $,
where $\Omega_\mu=\mu \Omega_0 - \mu \Delta$ and $\Omega_0 = c/L$ is the eigen-frequency of the ring resonator. Here, $\Delta$, $\alpha_\mu$, and $\phi_\mu$ denote the frequency detuning between the modulation and the ring resonator, the modulation amplitude, and the modulation phase, respectively. Under this periodic modulation, the eigenmodes of electromagnetic field becomes time-dependent. Thus, different modes are no longer orthogonal, resulting in effective couplings between two modes with different wave vectors.

In the rotation wave approximation, by adjusting the modulation signal of the phase modulator, we can arbitrary control the coupling strength between different lattice points.
To cut off the lattice size in the synthetic dimension, we introduce a small auxiliary resonator coupled to the large ring resonator~\cite{dutt_creating_2022}. Thus, the number of frequency levels $N$ can be determined by adjusting the size of the auxiliary resonator with a perimeter of $L_\mathrm{a}$. Then, the effective Hamiltonian of such a system in the synthetic dimension is generally written as ($\hbar=1$) 
\begin{equation}\label{eq4}
\begin{split}
{\hat{H}_0}=&- \sum_{\mu=1}^{M}   \sum_{n=1}^{N-\mu} K_\mu \frac{2\sqrt{\omega_n \omega_{n+\mu}}}{\left(\omega_n + \omega_{n+\mu}\right)} \left( e^{i \phi_\mu}\hat{a}_n^{\dagger}\hat{a}_{n+\mu}+{\rm h.c.} \right)\\
&- \sum_n^{N}  n\Delta \hat{a}_n^{\dagger}\hat{a}_n \\
&- \frac{g}{2} \sum_{nmpq} e^{i\Delta_kL}\mathrm{sinc}(\Delta_kL)  \hat{a}_n^{\dagger}\hat{a}_m^{\dagger}\hat{a}_p \hat{a}_q \delta_{n+m,p+q}.
\end{split}
\end{equation}
See details in Supplementary Materials (SM).
Here, $K_\mu = \alpha_\mu \Omega_0/2 \pi$ is the hopping matrix elements between levels $n$ and $n - \mu$ and we utilize three modulation modes, i.e. $M=3$.  The second term represents on-site potential for each frequency level.
The interaction strength $g=3\pi \hbar\omega_0^2 \chi^{(3)} /Vn_0^4 \epsilon_0$ represents the repulsion of photons at same levels, where $\hbar$ is the reduced Planck's constant, $\omega_0$ is the resonant frequency, $\chi^{(3)}$ is the third-order nonlinear susceptibility~\cite{strekalov_nonlinear_2016}, $V$ is the mode volume, $n_0$ is the refractive index. $\Delta_k = \beta_2(\Delta\omega_n^2 + \Delta\omega_m^2 -  \Delta\omega_p^2 - \Delta\omega_q^2)/2$, where $\Delta\omega_n = \omega_n - \omega_0 $ and the $\beta_2(\omega_0)$ is the GVD of the nonlinear media around $\omega_0$ with $\omega_0$ being the center frequency of synthetic frequency dimension. The creation and annihilation operators of the ring resonator satisfy the bosonic commutation relations.
In the regime with large GVD, some of those terms not satisfying the phase matching condition $\Delta_k = 0$ are suppressed due to the properties of $\mathrm{sinc}$ function. We focus on the large GVD regime since this regime exhibits rich quantum phases, in particular in systems with many photons. More details on the impact of GVD and the system size is discussed in SM. 


\begin{figure}
    \centering
    \includegraphics[width=\linewidth]{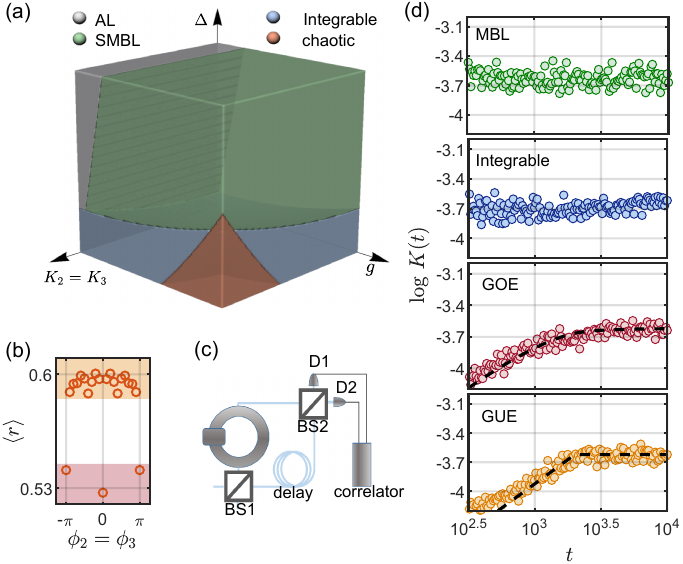}
  \caption{(a) Phase diagram of the ring resonator in the parameters of detunning $\Delta$,  long-range couplings $K_{2,3}$,  and  interaction strength $g$. (b) Statistical quantity $\langle r \rangle$ in the chaotic phase with different long-range coupling phases $\phi_2 = \phi_3$. The red area corresponds to GOE ensemble, and the yellow area corresponds to GUE ensemble. Device shown in (c) is used to measure SFF for obtaining phase diagram (a). Here, two beam splitters (BS1 and BS2) couple to optical fibers and two detectors connected to correlator. (d) The simulation of SFF for different parameter cases. For SMBL case, $g=1.5\Omega$, $K_2=K_3=1.25\Omega$ and $\Delta=3\Omega$. For integrable case, $g=1\Omega$, $K_2=K_3=0$, $\Delta=0$. For GOE and GUE cases, we have $g=1.5\Omega$, $K_2=K_3=1.25$, $\Delta=0$, while $\phi_2=\phi_3=0$ for GOE and $\phi_2=\phi_3=\pi/2$ for GUE, respectively. The black dashed lines are the theoretical estimation (details in SM). 
  }
  \label{fig:image2}
\end{figure}

{\bf Quantum phases---}The number of frequency levels $N$, the coupling strength $K_\mu$, the coupling phase $\phi_\mu$, the detunning $\Delta$, and the repulsion interaction $g$ are tunable. To characterize the quantum phases, we utilize a quantity of the level statistic $\langle r\rangle= \sum_{i=1}^{n-2}\frac{{\rm min}(s_{i+1},s_i)}{{\rm max}(s_{i+1},s_i)}/(n-2)$ in random matrix theory (RMT)~\cite{kriecherbauer_random_2001,PhysRevB.75.155111}, where the energy spacing is $ s_i = E_{i+1} - E_{i}$, and  $i$ is the eigenstate index. In general, $\langle r\rangle\approx 0.39$ stands for an integrable system. On the contrary, the system becomes quantum chaotic with or without the time-reversal symmetry corresponding to $\langle r\rangle\approx 0.53$ or $0.60$, dubbed as Gaussian orthogonal ensemble (GOE) or Gaussian unitary ensemble (GUE) according to RMT, respectively.  
By fixing the nearest-neighbor coupling strength at $K_1=1\Omega$ ($\Omega\ll \Omega_0$ serving as a reference value), we vary the next-nearest neighbor coupling strength $K_2$, the next-next-nearest neighbor coupling strength $K_3$, the interaction strength $g$, and the detunning $\Delta$ to construct a phase diagram for the large GVD regime, as shown in Fig.~\ref{fig:image2}(a). 

In the weak detunning regime, e.g. $\Delta < 0.2\Omega$, the behavior of system is jointly influenced by the long-range coupling strength $K_{2,3}$ and the interaction strength $g$. If either of them is weak, the system exhibits integrable behavior. Only when both the long-range coupling strength and the interaction strength are relatively large, the system enters the chaotic phase described by GOE. In addition, if the time-reversal symmetry is broken, altering the long-range coupling phase factor, the system transitions to chaotic phase described by GUE. As the strength of the detunning $\Delta$ gradually increasing, the system undergoes localization, exhibiting integrable behavior.

Furthermore, as shown in Fig.~\ref{fig:image2}(b), we propose a protocol to measure the spectral form factor (SFF), which is a quantity to statistically distinguish the degree of chaos. 
Its definition is given by
\begin{equation}\label{eq5}
K(t) =  \frac{1}{D^2} \overline{\Big[\sum_{\alpha \beta} C_{\alpha\beta} (t) e^{i \phi_{\alpha\beta}(t)}\Big]^2 },
\end{equation}
where $\alpha$ and $ \beta$ are the frequency indices of two photons. $D$ is the dimension of Hilbert space. The term $C_{\alpha\beta} (t) e^{i \phi_{\alpha\beta}(t)} $ can be obtained from the optical setup in Fig.~\ref{fig:image2}(c), by measuring the observable probability  $P^{n,m}_0 ={|\langle0|\hat{a}_{n,r}\hat{a}_{m,r}|\Psi\rangle|^2}/{(1+\delta_{nm})}$, $ P^0_{n,m}={|\langle0|\hat{a}_{n,t}\hat{a}_{m,t}|\Psi\rangle|^2}/{(1+\delta_{nm})} $, $ P^n_m={|\langle0|\hat{a}_{n,r}\hat{a}_{m,t}|\Psi\rangle|^2} $ through multiple repeated measurements (see details in SM). Here, subscripts `$r$' and `$t$' of $ \hat{a}_{n,r},\hat{a}_{n,t}$  represent the beam splitter reflection channel and transmission channel, respectively. 
According the result of multiple measurements, the result of the simulation based on this protocol is shown in Fig.~\ref{fig:image2}(d). 
$K(t)$ for the chaotic system exhibits unique traits, including a ramp before $t=t_H$ and a plateau for $t>t_H$. Here, $t_H$ is the Heisenberg time. Notably,  $K(t)$ curves for both the integrable and localized systems lack the ramp shape at its early time. Consequently, the degree of chaos of a system can be clearly deduced through the analysis of the SFF.

In the localized system, the interaction strength plays a crucial role in determining the evolution of the entanglement entropy. 
When $g$ approaches zero, the system becomes localized, which is characterized Bloch oscillations and the Wannier-Stark effect~\cite{wannier_wave_1960}.
Conversely, the system enters the realm of SMBL~\cite{van_nieuwenburg_bloch_2019}. In the SMBL phase, the system has a unique phenomenon that both the half-chain entanglement entropy of the system and the entanglement entropy of identical particles~\cite{amico_entanglement_2008,li_entanglement_2001} can continue to increase exponentially over time even that the imbalance has already been stabilized (see details in SM).
The photon state is ergodic in the Hilbert Space when the system is thermalized. While in the case of SMBL, the system is strongly localized and the entangled structure is stable. Therefore, it is possible to use a time-dependent detunning parameter to regulate the state of the photons, realizing entanglement control between photons.

\begin{figure}[ht]
 \centering
 \includegraphics[width=\linewidth]{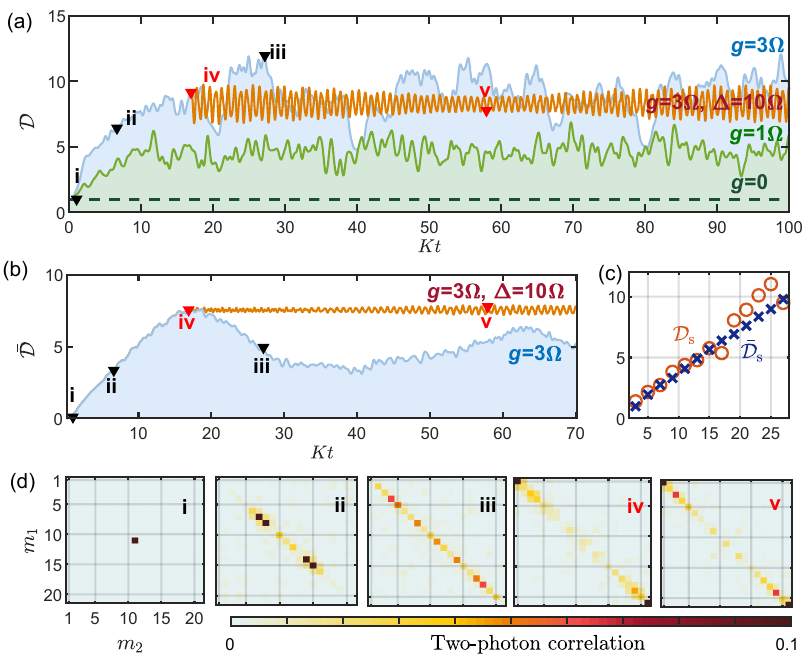}
 \caption{
  The evolution of $\mathcal{D}$ in (a) and $\bar{\mathcal{D}}$ in (b) of the two-photon state for different interaction strengths $g=0$, $1\Omega$, and $3\Omega$ without detunning $\Delta=0$. Here, the number of frequency levels is $N=21$ and the dispation rate is $\Gamma = 0.05\Omega$. The orange curve represents the evolution for $g=3\Omega$ and detunning $\Delta=10\Omega$, which is abruptly turned on at the time of point ``iv''. Then, $\mathcal{D}$ and $\bar{\mathcal{D}}$ slightly oscillate around their stable values. We define the stable values as $\mathcal{D}_\mathrm{s}$ and $\bar{\mathcal{D}_\mathrm{s}}$, which are plotted for different values of $N$ in (c). 
 Panel (d) shows the two-photon correlations at different points in (a) and (b). Other parameters are given by $K_1=1\Omega$, $K_2=K_3=0.3\Omega$, and $\phi_2=\phi_3=0$.
 }
  \label{fig:image3}
\end{figure}

{\bf Dynamics of entangled photons---}
Now, we study the ring resonator coupled to the input and output waveguides, as illustated in Fig.~\ref{fig:image1}(a). Here, we consider the environment bath coupling to both waveguides and the ring resonator and it results in the dynamics of the ring resonator as
\begin{equation}\label{eq6}
\begin{split}
\dot\rho_0(t) &= {-i}[\hat{H}_0,\rho_0]  \\
&+ \Gamma \sum_n \left( a_n \rho_0 a^{\dagger}_n - \frac{1}{2} a^{\dagger}_n a_n \rho_0 - \frac{1}{2} \rho_0 a^{\dagger}_n a_n \right),
\end{split}
\end{equation}
where $\rho_0$ is the reduced density matrix of the resonator and $\Gamma$ is the total decay rate of the ring resonator including both to the waveguides and to the environment bath.

To characterize the photon-photon entanglement in frequency, we introduce two quantities, degree of independent (DI) and weighted DI, respectively given by

\begin{equation}
\begin{aligned}
     \mathcal{D} = \frac{\sum_n P_{nn} }{\sum_n P^2_{n}},  \ \ \ \ \  \ \
     \overline{\mathcal{D}} = \frac{\sum_n 
     P_{nn} |\omega_n-\omega_0|}{\sum_n P_{nn} \Omega_0},
\end{aligned}
\end{equation}
where the probabilities of photon state is $P_{nm}={|\langle0|\hat{a}_n\hat{a}_m|\Psi\rangle|^2}/{(1+\delta_{nm})}$ and $P_n = P_{nn} +  \sum_{m\neq n}P_{nm}/2$. Here $(1+\delta_{nm})$ is normalization constant and $\omega_0$ is the central frequency in synthetic dimension chain.
The frequency distribution of two photons can be regarded as an independent event and thus $\mathcal{D} =1$ for the scenario of two independent photons, while $\mathcal{D}>1$ for the scenario of two entangled photons. The weighted DI $\overline{\mathcal{D}}$ depicts the similarity to the entangled state $\left(\left| \omega_{1} \omega_{1} \right\rangle + \left| \omega_{N} \omega_{N} \right\rangle\right)/\sqrt{2}$ with maximum frequency difference $\omega_N-\omega_1$.

Then, we employ Eq.~(\ref{eq6}) to obtain the time evolution of the two-photon state, where two input single photons are identical Gaussian wave packets with a standard deviation $\sigma = 0.01 \Omega_0$ in frequency domain and the central frequency of these two input photons are identical initially, as illustrated in Fig.~\ref{fig:image3}(a). 
The radiative decay rate of the ring resonator with a dissipation rate of $\Gamma = 0.05 \Omega$. We fix other parameters as $K_1=1\Omega$, $K_2=K_3=0.3\Omega$, and $\phi_2=\phi_3=0$. For scenarios with finite interaction strength, these parameters trigger the system into the quantum chaotic phase. The DI $\mathcal{D}$ dynamics exhibits rapid growth behavior at early times and then stop growth with severe fluctuations, as shown in Fig.~\ref{fig:image3}(a). In general, the growth of DI at early times is faster for the scenarios with stronger interaction strength $g$, while it stops at $\mathcal{D}=1$ for the non-interaction case.

For strong interaction case, say $g=3$, we set the initial state as a product state and it then evolves to  the entangled state $\left(\left| \omega_{1} \omega_{1} \right\rangle + \left| \omega_{N} \omega_{N} \right\rangle\right)/\sqrt{2}$. As shown in Fig.~\ref{fig:image3}(d), the two-photon correlation of the points ``i'', ``ii'', and ``iii'' in Fig.~\ref{fig:image3}(a), initially concentrates at central point, i.e. $m_1=m_2=11$ and then gradually splits along diagonal correlation. Finally, the correlation weight will spread out all diagonal correlations for $m_1=m_2$. The photon states can be approximately written as $\sum_n c_n |\omega_n\omega_n\rangle$ with $\sum_n c_n^\ast c_n=1$. 
The state at point ``iv'' in Fig.~\ref{fig:image3}(b), where $\overline{\mathcal{D}}$ attains its maximum, has been graphically represented in the Fig.~\ref{fig:image3}(d). The concentration of photon states at margins of the diagonal sites. At the time of point ``iv'' in Fig.~\ref{fig:image3}(b), $\overline{\mathcal{D}}$ arrives its maximum value and then we implement a finite detuning $\Delta=10$, resulting in a tilt potential in the synthetic frequency dimension. We find both quantities of $\mathcal{D}$ and $\bar{\mathcal{D}}$ become stable only with slight oscillation, reminiscent of SMBL in quantum many-body models in real space~\cite{schulz_stark_2019}. We also check the stable DI ${\mathcal{D}}_\mathrm{s}$ and the stable weighted DI $\overline{\mathcal{D}}_\mathrm{s}$ for different numbers of frequency levels $N$. As shown in Fig.~\ref{fig:image3}(c). We find that $\overline{\mathcal{D}}_\mathrm{s}$ can be linearly tuned by the frequency mode number $N$, which is determined by the small auxiliary resonator in Fig.~\ref{fig:image1}(a).
In addition, We find that the dissipation of photons does not affect the results of the detected two-photon states.

\begin{figure}[ht]
\centering
\includegraphics[width=0.75\linewidth]{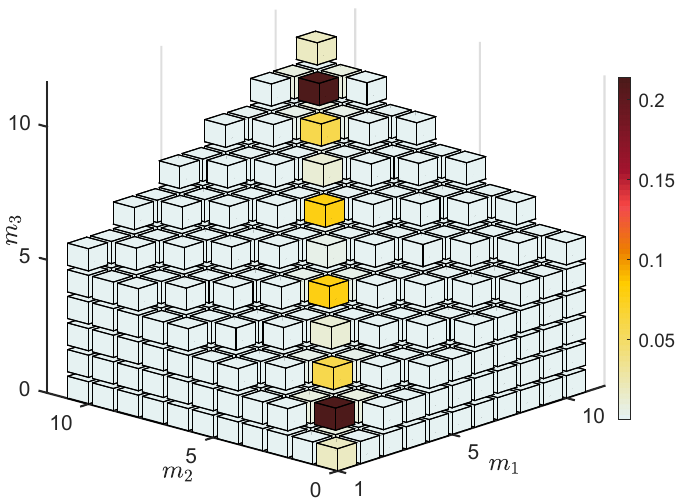}
\caption{Three-photon correlation for $g=3 \Omega$, $\Delta=3 \Omega$, and $N = 11$. $x$, $y$ and $z$ axes represent the indices of frequency levels.}
\label{fig:image4}
\end{figure}

Finally, we notice that this method to generate entangled-photon states can be extended to the scenarios of more photons. As shown in Fig.~\ref{fig:image4}, we plot the three-photon entangled states.

{\bf Discussion--}
This work presents the first systematic investigation of few-body quantum model in optical synthetic frequency dimension. This system has the potential to exhibit a rich variety of quantum phases. By coupling it to input and output waveguides, we can dynamically control these phases and generate entangled photon states.
Previously, some traditional methods exist for generating various types of entangled photon states, such as entanglement between frequencies and orbital angular momenta (OAM)~\cite{PhysRevLett.122.083903}, as well as frequency-bin entangled states generated through the spontaneous parametric down-conversion~\cite{frequency-bin,PRA-frequency-bin}. 
Unlike the co-frequency state discussed here, frequency-bin entangled photons occupy distinct spatial positions, making them distinguishable. Furthermore, their entanglement arises from second-order nonlinearities. In contrast, the co-frequency state describes the entire two-photon system. It relies on third-order nonlinearities to generate two high-frequency photons.
In addition, the third-order nonlinear materials are typically valid in the range of 120~THz to 700~THz~\cite{Rid}. 
This allows us to construct a frequency domain spanning $\Delta_f = 580$~THz, corresponding to the light from the infrared to the violet spectrum. The number of levels in the model is determined by the perimeter of the ring resonant $N_{\rm lp} = \Delta_f L/c$. Thus, This approach enables the generation of entangled states between $|\omega_N\cdots \omega_N\rangle$ and $|\omega_1\cdots \omega_1\rangle$, even when they have a significant difference in frequency. 
 
Furthermore, the synthetic frequency dimension could transcends the limitations of traditional 3+1D space~\cite{Cheng2023,Dutt2020,jukic2013four,lustig2019photonic,boada2012quantum} and thus it enables the exploration of few-body and many-body phenomena in higher dimensions. The additional presence of atom-light interaction allows for the investigation of a wide range of entanglement phenomena through its ability to generate them~\cite{Javid2023}. 
This will lead to an expansion in the complexity of the many-body phenomena beyond traditional spatial limitations that can be studied on synthetic dimensions.

\vspace{+0.2cm}

\noindent{\bf Acknowledgement}\\
This work was supported by National Natural Science Foundation of China (grants Nos. 12375021, 12122407, and 12204304) and National Key R\&D Program of China (grants No. 2022YFA1404203 and 2023YFA1407200). L.Y. also thanks the sponsorship from the Yangyang Development Fund.
%


\clearpage

\begin{onecolumngrid} %

\beginsupplement

\begin{center}
{\bf {\large Supplementary Materials}}
\end{center}

%

\section{Derivation of the effective Hamiltonian} \label{S1}

We construct a micro-nano optical system, and obtain its quantum Hamiltonian through the second quantization of electromagnetic field. The micro-nano optical device consists of a ring resonant and a periodic electro-optical modulation $\phi(t)$:
\begin{equation}\label{eqs1_1}
\begin{aligned}
\phi(t) = \sum_{\mu=1}^{M} \alpha_{\mu}\left[ e^{i\left( \Omega_{\mu} t+\phi_\mu \right) } + e^{-i\left( \Omega_{\mu} t+\phi_\mu \right) }\right] ,
\end{aligned}
\end{equation}
where $\Omega_\mu=\mu(\Omega_0 - \Delta)$ with $\Omega_0 = c/L$. $\Delta$ is the detuning between frequency of modulator and eigenfrequency of the ring resonator. $\alpha_{1,2,3}$ are modulation amplitudes. Due to the periodic electro-optical modulation, the electromagnetic field operators in the system are as follows:
\begin{equation}\label{eqs1_2}
\begin{split}
\hat{E}(z,t) &=i \sum_{n} \sqrt{ \frac{h \omega_n}{2 V \epsilon_r \epsilon_0} } \left\{ \hat{a}_n  {\rm sin}\left[\left(k_n + \frac{\phi (t)}{L}\right) z\right]-\hat{a}_n^\dagger   {\rm sin}\left[\left(k_n + \frac{\phi (t)}{L}\right) z\right]\right\},\\
\hat{B}(z,t) &= \sum_{n} \frac{1}{c}\sqrt{ \frac{h \omega_n}{2 V \epsilon_r \epsilon_0} } \left\{ \hat{a}_n  {\rm sin}\left[\left(k_n + \frac{\phi (t)}{L}\right) z\right]+\hat{a}_n^\dagger   {\rm sin}\left[\left(k_n + \frac{\phi (t)}{L}\right) z \right]\right\},
\end{split}
\end{equation}
where $\hat{a}_n^\dagger$ ($\hat{a}_n$) is creation(annihilation) operator of the $n$th mode. $k_n= {2 n\pi }/{L}$, where $L$ is the perimeter of the ring resonator, $n$ is an integer number. The $z$-direction is the angular direction of the ring resonator. Because of the periodic electro-optical modulation, the modes of electric field and magnetic field become time-dependent and different modes are no longer orthogonal to each other all the time, which leads to the coupling between two modes. The inner products between the same modes and different modes are given by
\begin{equation}\label{eqs1_3}
\begin{aligned}
\frac{2}{L}\int_0^L {\rm sin}\left[\left(k_j + \frac{\phi (t)}{L}\right) z\right]&{\rm sin}\left[\left(k_j + \frac{\phi (t)}{L}\right) z\right] dz \approx 1 ,\\
\frac{2}{L}\int_0^L {\rm sin}\left[\left(k_i + \frac{\phi (t)}{L}\right) z\right]&{\rm sin}\left[\left(k_j + \frac{\phi (t)}{L}\right) z\right] dz 
 = -\frac{{\rm sin}\left[\frac{(i+j)2\pi+2\phi (t)}{L} z\right]|_0^L}{(i+j)2\pi+2\phi (t)} \approx - \frac{\phi(t)}{(i+j)\pi} .
\end{aligned}
\end{equation}
The Hamiltonian of the electromagnetic field in the system is written as
\begin{equation}\label{eqs1_4}
\mathcal{H}=\frac{1}{2}\int dv  \left(\epsilon_r \epsilon_0 \hat{E}^2 + \frac{1}{\mu_r\mu_0}\hat{B}^2\right).
\end{equation}
Leveraging Eq. ( \ref{eqs1_3} ) and substituting Eq. ( \ref{eqs1_2} ) into Eq. ( \ref{eqs1_4}), we have the following result:
\begin{equation}\label{eqs1_5}
\begin{aligned}
\hat{H}=&- \sum_n^{N}  \omega_m \hat{a}_n^{\dagger}\hat{a}_n 
- \sum_{n \neq m} ^{N} \frac{c \sqrt{\omega_n \omega_m} \phi(t)}{(\omega_n + \omega_m) \pi L} \hat{a}_n^{\dagger}\hat{a}_{m} ,
\end{aligned}
\end{equation}
where $\omega_n$ is the frequency of the $n$th mode.
After taking the frame rotation $\hat{a}_m \rightarrow \hat{a}_m e^{i (\omega_m-\Delta) t}$ and the rotating wave approximation, we have
\begin{equation}\label{eqs1_6}
\begin{aligned}
\hat{H}=&- \sum_m^{N} m \Delta \hat{a}_m^{\dagger}\hat{a}_m - \sum_{\mu=1}^{M}   \sum_{n=1}^{N-\mu} K_\mu \frac{2\sqrt{\omega_n \omega_{n+\mu}}}{\left(\omega_n + \omega_{n+\mu}\right)} \left( e^{i \phi_\mu}\hat{a}_n^{\dagger}\hat{a}_{n+\mu}+{\rm h.c.} \right).
\end{aligned}
\end{equation}
where $M = 3$ and $K_1 = {\alpha_1 \Omega_0}/{2 \pi}$, $K_2={\alpha_2 \Omega_0}/{2 \pi}$, $K_3={\alpha_3 \Omega_0}/{2 \pi}$. The period boundary condition of the ring resonator creates a series of modes and this kind of boundary condition constructs a synthetic frequency dimension, whose spacing between two nearest-neighbored frequency levels depends on the ring resonator perimeter $L$. To constitute an extended synthetic frequency``chain'', the electro-optic modulator couples different frequency modes .

If the ring resonator is made by nonlinear materials with a finite third-order coefficient, the Hamiltonian describing such a nonlinear system requires an additional term given by
\begin{equation}\label{eqs1_7}
\begin{aligned}
\hat{H}_{\rm int} = \int d{\rm V} \sum_{nmpq} \chi^{(3)} E_n E_m E_p E_q,
\end{aligned}
\end{equation}
where $\chi^{(3)}$ is the third-order nonlinear coefficient and the subscription of $E_n$ is frequency indices. The media in the resonant ring is divided into two parts, which have the same third-order nonlinear coefficient but have opposite group velocity dispersions (GVD). The third-order nonlinear media can effectively induce the interaction between photons.
The Hamiltonian consists of two parts:
\begin{equation}\label{eqs1_8}
\begin{aligned}
\hat{H}_{\rm int} = {S}\int_0^{L} dz \sum_{nmpq} \chi^{(3)} E_n E_m E_p E_q= S\int_0^{L/2} dz  \sum_{nmpq} \chi^{(3)} E_n E_m E_p E_q +  S\int_{L/2}^L dz  \sum_{nmpq} \chi^{(3)} E_n E_m E_p E_q,
\end{aligned}
\end{equation}
where $S$ is the cross-sectional area of the ring resonator. The quantization form of the electric field is written as
\begin{equation}\label{eqs1_9}
\begin{aligned}
\hat{E}(z,t)=i \sum_{n} \sqrt{ \frac{h \omega_n}{2 V \epsilon \epsilon_0} }  \left\{ \hat{a}_n E_n(z)-\hat{a}_n^\dagger E^*_n(z) \right\}.\\
\end{aligned}
\end{equation}
The spatial function $E_n(z)$ is a segmentation function as
\begin{equation}\label{eqs1_10}
E_n(z) = \left\{
        \begin{aligned}
	&e^{i\left[k_n+(\omega_n-\omega_0)/v_g + \beta_2(\omega_n-\omega_0)^2\right]z}, \quad 0<z<L/2\\
	&e^{i\left[k_n+(\omega_n-\omega_0)/v_g -\beta_2(\omega_n-\omega_0)^2\right]z}, \quad L/2<z<L\\
 \end{aligned}
	\right.
\end{equation}
where $v_g = (d\beta / d\omega)^{-1}$ is the group velocity and $\beta_2(\omega_0) = d^2\beta(\omega_0) / d\omega^2$ is the GVD around $\omega_0$. $\beta$ is the wavevector and $\beta_2(\omega_0)$ is the second-order Taylor expansion coefficient of $\beta$ at $\omega = \omega_0$.
Then, substituting Eq.~(\ref{eqs1_9}) into Eq.~( \ref{eqs1_8}), we have a definite integral as
\begin{equation}\label{eqs1_12}
\begin{aligned}
&S \chi^{(3)} \int_0^{L/2} dz E_n E_m E_p E_q \\
 = & S \chi^{(3)} \frac{ \sqrt{\omega_n \omega_m \omega_p \omega_q} }{4 V^2 \epsilon_r^2 \epsilon_0^2} \binom{2}{4} \int_0^{L/2} dz e^{i(k_n+\beta_2\Delta \omega_n^2)z} e^{i(k_m+\beta_2\Delta\omega_m^2)z}e^{-i(k_p+\beta_2\Delta\omega_p^2)z} e^{-i(k_q+\beta_2\Delta\omega_q^2)z}\hat{a}^{\dagger}_n \hat{a}^{\dagger}_m \hat{a}_p \hat{a}_q,
\end{aligned}
\end{equation}
where $\Delta\omega_n = \omega_n - \omega_0$.
The term ($\hat{a}^{\dagger}_n \hat{a}^{\dagger}_m \hat{a}_p \hat{a}_q$) describes the four-wave-mixing process caused by third-order nonlinear susceptibility. These high-order terms satisfy $n+m=p+q$ and their strengths are given by $e^{i\Delta_kL}\mathrm{sinc}(\Delta_kL)$, where $\Delta_k = \beta_2 (\Delta\omega_n^2 + \Delta\omega_m^2 -  \Delta\omega_p^2 - \Delta\omega_q^2)/2 $. Thus, the effective Hamiltonian is given by
\begin{equation}\label{eqs1_13}
\begin{aligned}
\hat{H}_0=&- \sum_n^{N} n \Delta \hat{a}_n^{\dagger}\hat{a}_n - \sum_{\mu=1}^{M}   \sum_{n=1}^{N-\mu} K_\mu \frac{2\sqrt{\omega_n \omega_{n+\mu}}}{\left(\omega_n + \omega_{n+\mu}\right)} \left( e^{i \phi_\mu}\hat{a}_n^{\dagger}\hat{a}_{n+\mu}+{\rm h.c.} \right)
 - g/2 \sum_{nmpq} e^{i\Delta_kL}\mathrm{sinc}(\Delta_kL)  \hat{a}_n^{\dagger}\hat{a}_m^{\dagger}\hat{a}_p \hat{a}_q \delta_{n+m,p+q},
\end{aligned}
\end{equation}
where $g=3h\omega_0^2 \chi^{(3)} /Vn_0^4 \epsilon_0$ is the interaction strength. The creation (annihilation) operators corresponding to the modes in the ring resonant satisfy the Bosonic commutation relations:
\begin{equation}\label{eqs1_14}
    \begin{aligned}
 & [\hat{a}_n,\hat{a}_m^\dagger] = \delta_{n,m} \quad,\quad[\hat{a}_n^\dagger,\hat{a}_m^\dagger] = 0 \quad,\quad [\hat{a}_n,\hat{a}_m] = 0 .\\
    \end{aligned}
\end{equation}

Importantly, $\beta_2$ can regulate the interaction strength. Most terms are negligible due to the rapid decay of the $\mathrm{sinc}$ function (${\rm sinc}(x) = {\sin (x)}/{x} $) away from $x=0$, in particular when the value of $\beta_2$ is large. In specific, the SPM term ($\hat{a}_m^{\dagger}\hat{a}_m^{\dagger}\hat{a}_m \hat{a}_m$) and the XPM term ($\hat{a}_n^{\dagger}\hat{a}_m^{\dagger}\hat{a}_n \hat{a}_m$) satisfy  $\Delta\omega_n^2 + \Delta\omega_m^2 -  \Delta\omega_p^2 - \Delta\omega_q^2 = 0$. Therefore, the strength for these terms can be approximated by $g/2$. When the GVD ($\beta_2$) is large, the interaction Hamiltonian becomes:
\begin{equation}\label{eqs1_15}
\begin{aligned}
\hat{H}_{\rm int}=
&-\frac{g}{2} \sum_m \hat{a}_m^{\dagger}\hat{a}_m^{\dagger}\hat{a}_m \hat{a}_m-g \sum_{m \neq n} \hat{a}_m^{\dagger}\hat{a}_m \hat{a}_n^{\dagger}\hat{a}_n, \\
\end{aligned}
\end{equation}
Due to the U(1) symmetry of the Hamiltonian, the interaction Hamiltonian can be further transformed to the form only with local interactions, as expressed by
\begin{equation}\label{eqs1_16}
\begin{aligned}
\hat{H}_{\rm int}^{\prime}=&-\frac{g}{2} \sum_m \hat{a}_m^{\dagger}\hat{a}_m^{\dagger}\hat{a}_m \hat{a}_m-g \sum_{m} \hat{a}_m^{\dagger}\hat{a}_m(N_p - \hat{a}_m^{\dagger}\hat{a}_m )\\
=&\frac{g}{2} \sum_m \hat{a}_m^{\dagger}\hat{a}_m^{\dagger}\hat{a}_m \hat{a}_m-g N_p^2,
\end{aligned}
\end{equation}
where $N_p$ is the photon number. The effective Hamiltonian can be written as
\begin{equation}
\begin{aligned}
\hat{H}_0^{\prime}=&-h \sum_n^{N} n \Delta \hat{a}_n^{\dagger}\hat{a}_n - \sum_{\mu=1}^{M}   \sum_{n=1}^{N-\mu} K_\mu \frac{2\sqrt{\omega_n \omega_{n+\mu}}}{\left(\omega_n + \omega_{n+\mu}\right)} \left( e^{i \phi_\mu}\hat{a}_n^{\dagger}\hat{a}_{n+\mu}+{\rm h.c.} \right) +\frac{g}{2} \sum_m \hat{a}_m^{\dagger}\hat{a}_m^{\dagger}\hat{a}_m \hat{a}_m.
\end{aligned}
\end{equation}

\section{quantum phase in different parameter condition} \label{QP}

\begin{figure}
  \centering
    \includegraphics[width=1\linewidth]{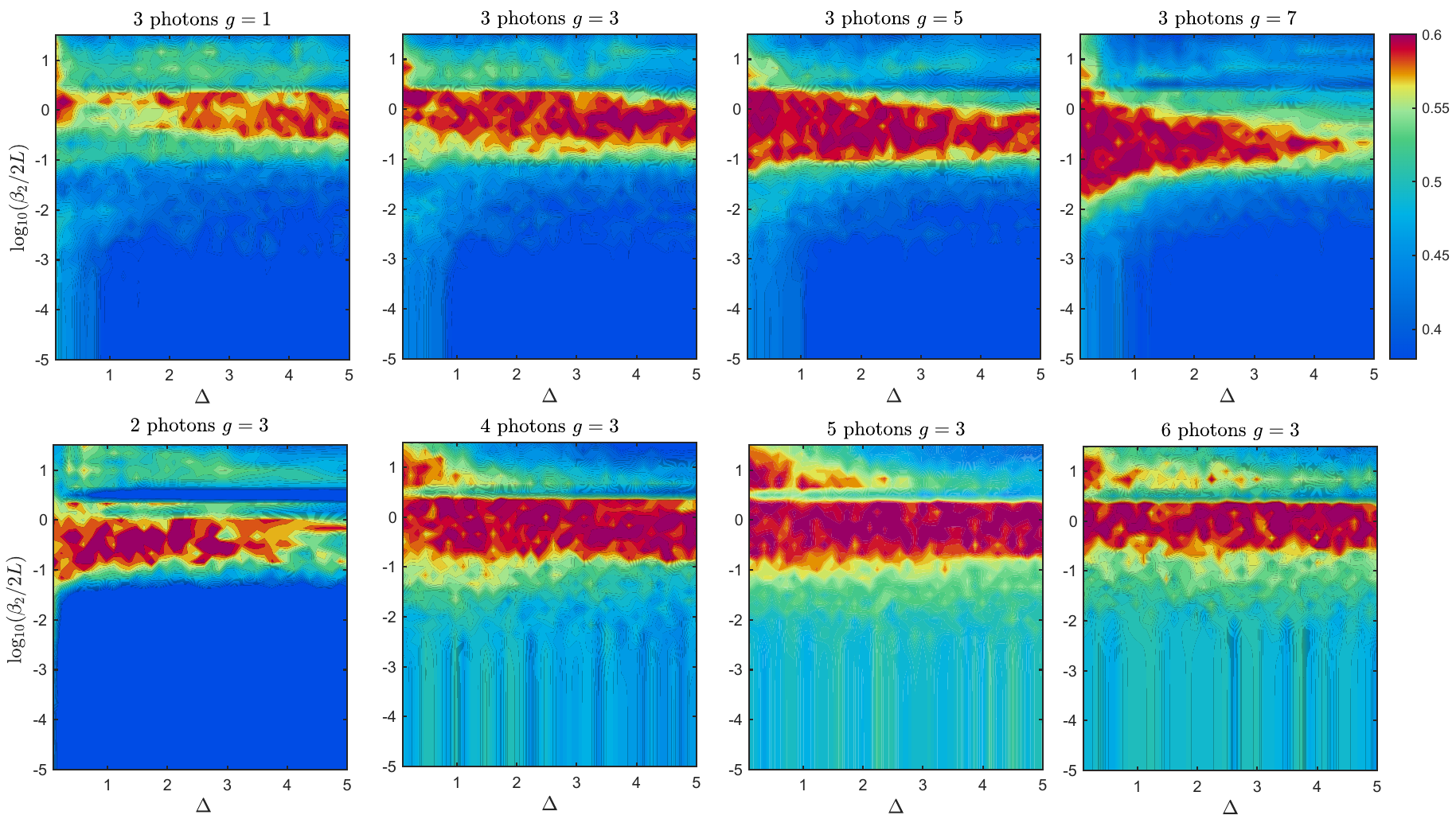}
  \caption{$\langle r\rangle$ statistics for different parameters and different photon numbers. Here, we fix $K_1=1$ and $K_2 = K_3 =0$.}
  \label{fig:S1}
\end{figure}

\subsection{Quantum phases for varying parameters}
Here, we discuss the impact of some parameters of the system. Firstly, we focus on how the parameter $\beta_2$ affects the chaotic degree of the system. In Fig.~\ref{fig:S1}, we know that the quantum system is approximately integrable when $\beta_2$ is very large or very small and $\Delta$ is large. Conversely, for intermediate values of $\beta_2$, the system exhibits chaotic behavior. Remarkably, this chaotic behavior caused by $\beta_2$ are robust for all values of $\Delta$. 
However, increasing the number of particles within the system also can promote a tendency towards chaos. Notably, the system undergoes a transition from chaotic to integrable behaviors when $\beta_2$ is large and $g$ falls within a specific range. Therefore, we choose the $\beta_2$ in this range in the main text.
We observe a region of integrable parameters in Fig. \ref{fig:S1} around $\beta_2/2L=\pi/2$. This arises due to the nature of the $\mathrm{sinc}$ function, which equalizes the strengths of various high-order terms.

We note that the purpose of two nonlinear media with opposite GVD and equal lengths is to satisfy the GVD matching condition: $L_1\beta_2(1) = L_2\beta_2(2)$, where $L_{1,2}$ are the lengths of two nonlinear media and $\beta_2(1,2)$ are the $\beta_2$ of two nonlinear media. It ensure that the total GVD of the ring resonant is zero. This condition introduces a quadratic tilt potential into the system. This tilt potential cannot be eliminated by simply adjusting the detuning between the electro-optic modulation and the eigenfrequency of ring resonator. Consequently, if the tilt potential becomes too large, manipulating the detuning will no longer be sufficient to achieve the desired phase transition from thermalization to localization.

\begin{figure}
  \centering
    \includegraphics[width=1\linewidth]{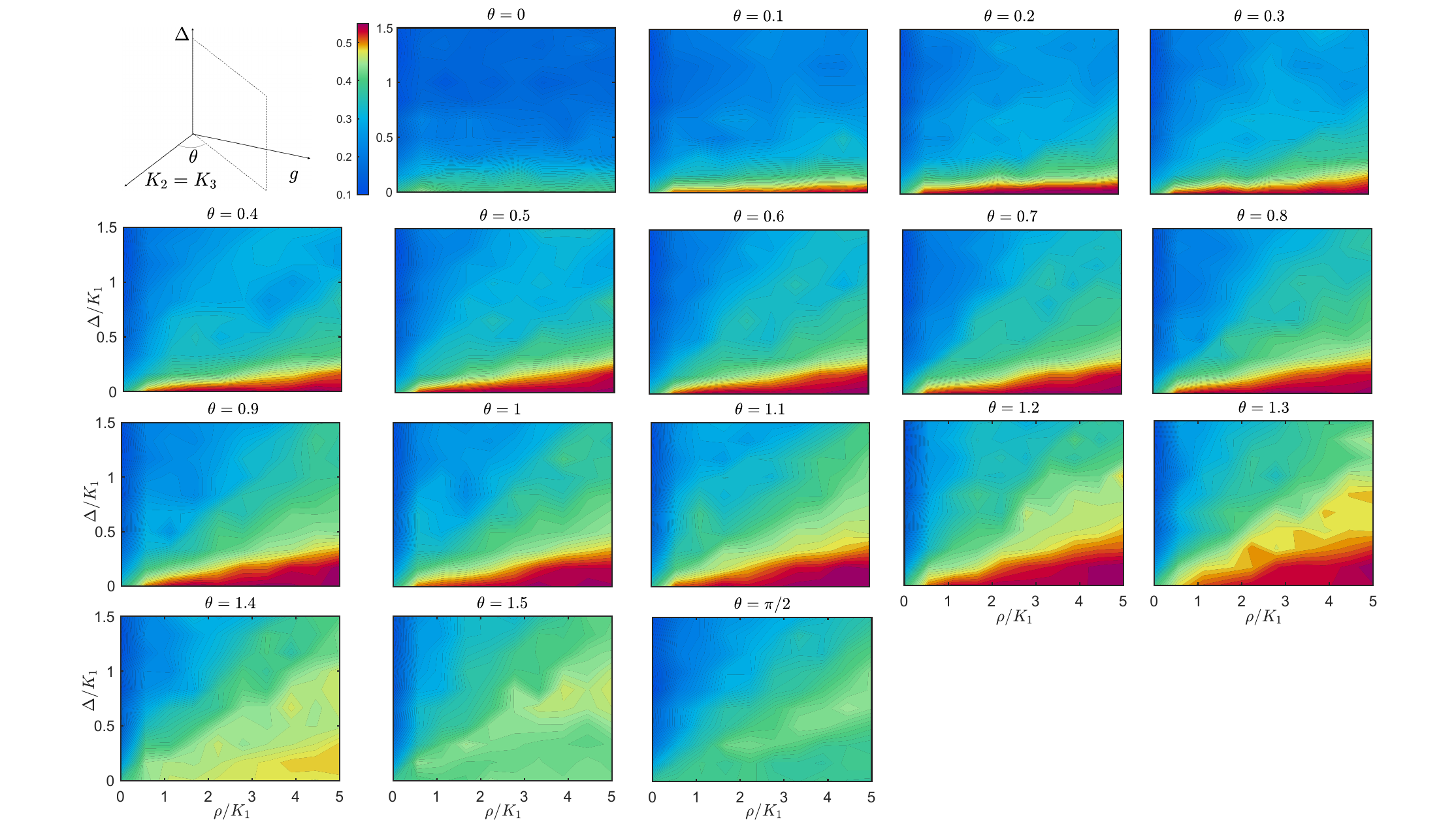}
  \caption{$\langle r\rangle$ for different parameter section planes. The section planes we selected are perpendicular to the $\Delta=0$ plane with an angle $\theta$ to the $g=0$ plane. Here, $\theta$ is set from $0$ to $\pi/2$.}
  \label{fig:S3}
\end{figure}

Fig.~\ref{fig:image2}(a) in the main text is a phase diagram of quantum system under different parameters, based on the numerical simulation results in the Fig.~\ref{fig:S3}. In order to understand the chaotic properties of the system in the space with three parameters, we take $17$ cross sections and calculate $\langle r\rangle$ on them. When the system is in chaos, the $\langle r\rangle$ approaches $0.53$. When the system is integrable and SMBL, the $\langle r\rangle$ approaches $0.39$.

\subsection{Localized phase in strong detunning regime}

Here, we illustrate the localized behaviors across a range of interaction strengths $g$, from weak ($g = 0.001$) to strong ($g = 10$). we set $\beta_2/2L=50$, which effectively cancels out the contributions of other high-order terms in Eq.~(\ref{eq4}). When the interaction is weak, as shown in the Fig.~\ref{fig:S_space}, in the Bosonic system we can observe the typical phenomenon of localization: even the imbalance is stabilized as the time evolution, the half-chain entanglement entropy of the system continues to grow linearly with time on an exponential scale~\cite{schulz_stark_2019}. Here, the imbalance is defined as $I(t) = \sum_i \langle \Psi(0) | n_i |\Psi(0) \rangle \langle \Psi(t) | n_i | \Psi(t) \rangle$, where $n_i = a_i^\dagger a_i$ represents photons number operator and the half-chain entanglement entropy  $S_{\mathrm{sp}} = {\rm Tr}\left(\rho _A{\rm log}\left( \rho_A \right) \right)$ is calculated by equally cut the system into two parts. 

Here, we further investigate the evolution of a different type of entanglement entropy that characterizes the degree of entanglement between frequencies of two identical photons. The photon entanglement entropy $S_{\mathrm{ph}}$ defined as the von Neumann entropy between two photons, which can be calculate by performing the Schmidt decomposition of the wave function~\cite{amico_entanglement_2008,li_entanglement_2001}: $|\Psi(1,2)\rangle = \sum_{i,j}C_{i,j}|i\rangle \otimes|j\rangle$. It should satisfy: $C_{i,j}=C_{j,i}$ and $\sum_{i,j}|C_{i,j}|^2=1$.  It exhibits a phenomenon similar to $S_{\mathrm{sp}}$. We simulate the evolution of a special case where two photons begin at the same frequency level, which has well-defined entanglement entropy~\cite{li_entanglement_2001}. Figure~\ref{fig:S_photo}(a) shows how the entanglement entropy and imbalance, denoted by $I(t)$, of entangled photons evolve with varying interaction strengths $g$. When the interaction strength is less than a certain value, even if the imbalance of the photons has stabilized, the $S_{\mathrm{sp}}$ continues to rise and eventually reaches a stable value.

We find that the stable values of the photon entanglement entropy, $S_{\mathrm{phs}}$, differ at various parameter conditions, as illustrated in Fig.~\ref{fig:S_photo}(a). We note that $S_{\mathrm{phs}}$ is influenced by both the interaction strength $g$ and the detunning $\Delta$, as shown in Fig.~\ref{fig:S_photo}(b). Notably, $S_{\mathrm{phs}}$ is significantly larger at the rectilinear area with an integer slope. In these parameters, the initial state will be in a larger Hilbert space with the similar energy. Despite the absence of direct coupling between these states, the initial state can still evolve across all of them due to the quantum tunneling effect~\cite{scherg_observing_2021}. As a result, $S_{\mathrm{phs}}$ will be larger when the ratio of the interaction strength to the tilting strength is an integer. Consequently, $S_{\mathrm{phs}}$ is expected to be higher when the ratio of interaction strength $g$ to tilting strength $\Delta$ is an integer.

\begin{figure}
  \centering
    \includegraphics[width=0.4\linewidth]{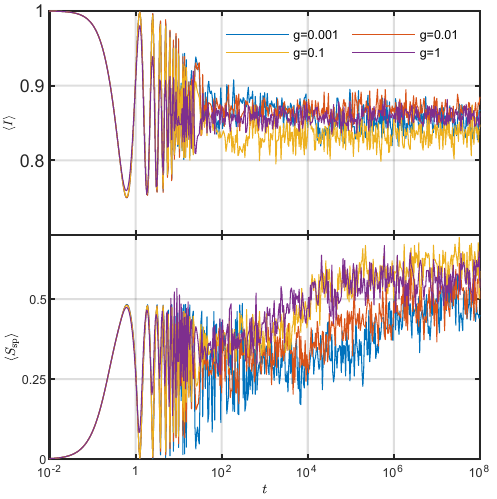}
  \caption{Evolution of the local observable imbalance, $I(t)$, and the half-chain entanglement entropy, $S_{\mathrm{sp}}$, for a half-fill chain and $N=10$. For all the evolution in the figure, we fix $\Delta = 5$ and $K_2 = K_3 = 0.3$. The initial state is $|0101\dots\rangle$. We perform ensemble averaging of the evolution over 10 different disorder realizations.}
  \label{fig:S_space}
\end{figure}

\begin{figure}
  \centering
    \includegraphics[width=0.85\linewidth]{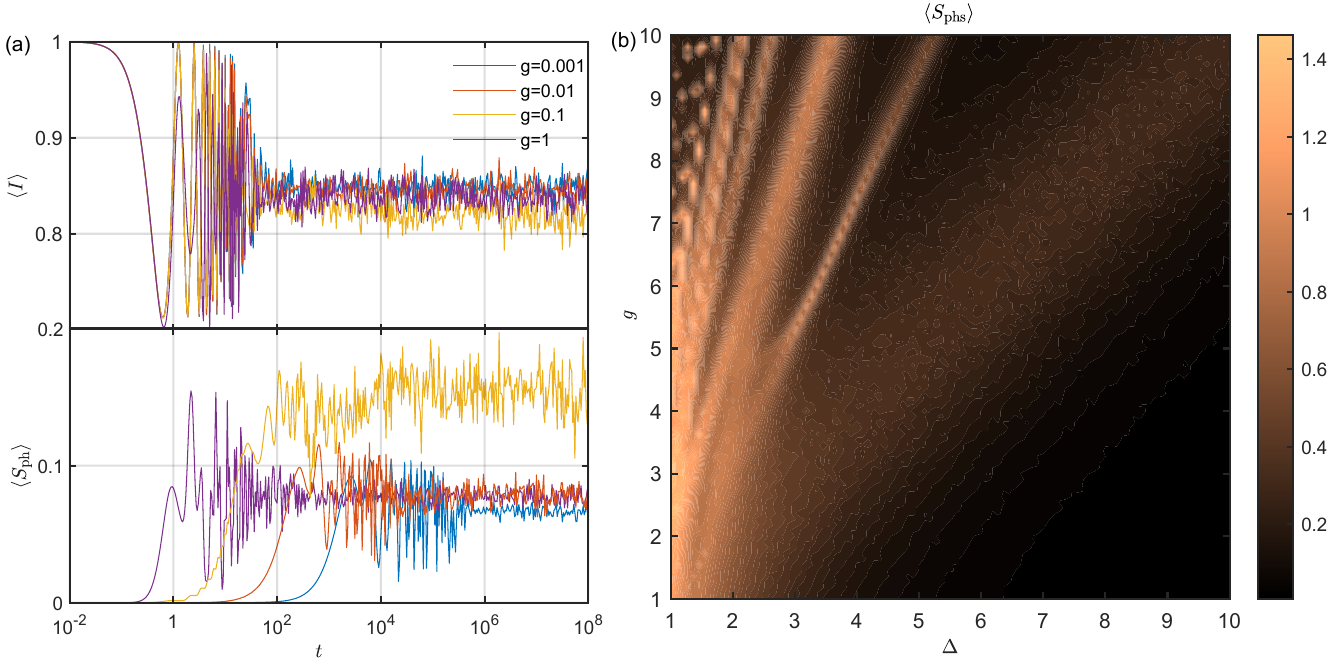}
  \caption{(a) Evolution of the local observable imbalance, $I(t)$, and the photon entanglement entropy, $S_{\mathrm{ph}}$, for two photons and $N=51$. Throughout all the evolution depicted in the figure, we fix $\Delta=5$ and $K_2 = K_3=0.3$, respectively. For different parameter conditions, we select the 43 frequency levels in the middle of the synthetic frequency chain as the initial positions of the two photons for averaging. (b) The stable photon entanglement entropy on average for varying values of $g$ and $\Delta$, while fixing $K_2 = K_3 = 0.3$ and $N=51$.}
  \label{fig:S_photo}
\end{figure}

\section{Protocol of spectral form Factor} \label{SFF}

The statistical properties of the energy spectrum and  $\langle r \rangle$ statistic value can be used to quantify the degree of chaos in a system. However, in general, they are hard to be directly measured. Here, we propose a protocol in such a system to measure the SFF. The SFF is defined by
\begin{equation}
\begin{aligned}
K(t) = \frac{1}{D^2} \overline{{\rm Tr}[\hat{T}(t)] {\rm Tr} [\hat{T}(t)^\dagger]},
\end{aligned}
\end{equation}
where $\hat{T}(t)$ denotes the time evolution operator. Its trace, denoted by $\hat{T}(t)$, is:
\begin{equation}
\begin{aligned}
 {\rm Tr} [\hat{T}(t)]  = {\rm Tr} [e^{i\hat{H}t/h}] = \sum_n \langle\Psi_n|e^{i\hat{H}t/h}|\Psi_n\rangle=\sum_{\alpha \beta} \langle \omega_\alpha \omega_\beta | \hat{T}(t) | \omega_\alpha \omega_\beta \rangle = \sum_{\alpha \beta} C_{\alpha\beta} (t) e^{i \phi_{\alpha\beta}(t)}.
\end{aligned}
\end{equation}
We can choose the product states of two photon frequencies, $| \omega_\alpha \omega_\beta \rangle$ as the complete basis $|\Psi_n\rangle$. Here, $\omega_\alpha$ and $\omega_\beta$ represent the individual frequencies of the two photons. Both $C_{\alpha\beta}$ and $\phi_{\alpha\beta}$ are real numbers. 
The Spectral Form Factor (SFF), denoted by K(t), for a two-photon interaction system can be expressed as:
\begin{equation}
\begin{aligned}
K(t) = \frac{1}{D^2} \overline{\Big|\sum_{\alpha \beta} C_{\alpha\beta} (t) e^{i \phi_{\alpha\beta}(t)}\Big|^2 }.
\end{aligned}
\end{equation}
The SFF function $K(t)$ can be calculated by
$C_{\alpha\beta} (t) e^{i \phi_{\alpha\beta}(t)}$  at different times, $t$. These complex quantities containing information about the system at time $t$. In this regard, we will present a method for measuring these quantities using two beam splitters. The experimental setup is detailed in Fig.~\ref{fig:image2}(c). Two photons with frequencies $\omega_\alpha$ and $\omega_\beta$ are first combined on a beam splitter (BS1). The reflected light is coupled to the ring resonator, while the transmitted light is directed to an optical delay line. The two outputs are then recombined at BS2. Then, probabilities such as $P\left( | \omega_\alpha, \omega_\beta\rangle_{r}|0\rangle_t \right)$, $P\left( |0\rangle_{r}| \omega_\alpha, \omega_\beta\rangle_t \right)$, $P\left(| \alpha\rangle_{r}| \omega_\beta\rangle_t \right)$, and $P\left( | \omega_\beta\rangle_{r}| \alpha\rangle_t \right)$ can be obtained.
 
Now, we derive the expressions for these probabilities. The quantum representation of beam splitters is:
\begin{equation}
\begin{aligned}
& \hat{a}_{in}^\dagger =\frac{1}{\sqrt{2}}\left(\hat{a}_r^\dagger  + \hat{a}_t^\dagger  \right) , \\
& \hat{a}_{u}^\dagger =\frac{1}{\sqrt{2}}\left(\hat{a}_r^\dagger  - \hat{a}_t^\dagger  \right) .
\end{aligned}
\end{equation}
Here, the subscripts $u$, $in$, $r$, and $t$ denote two input ports (denoted by u and in), reflection output port (r), and transmission output port (t) of the beam splitters, respectively. We begin by considering a single-photon state $|\omega_\alpha\rangle$, as the incident state. In this case, the output photon state can be mathematically expressed:
\begin{equation}
\begin{aligned}
|\omega_\alpha\rangle_{in} |0\rangle_u& = \hat{a}_{in}^\dagger (\omega_\alpha) |0\rangle_{in} |0\rangle_u\\
& =\frac{1}{\sqrt{2}}\left(\hat{a}_r^\dagger (\omega_\alpha) + \hat{a}_t^\dagger (\omega_\alpha) \right)  |0\rangle_{r} |0\rangle_t\\
& =\frac{1}{\sqrt{2}}\left( |\omega_\alpha\rangle_r |0\rangle_t + |0\rangle_r |\omega_\alpha\rangle_t \right) .
\end{aligned}
\end{equation}
After evolving, the initial state $|\omega_\alpha\rangle_r |0\rangle_t$ transforms into $\sum_\mu C_\mu e^{i \phi_\mu } |\omega_\mu \rangle_r |0\rangle_t$. These photon state then recombine at BS2:
\begin{equation}
\begin{aligned}
&\frac{1}{\sqrt{2}}\left( \sum_\mu C_\mu e^{i \phi_\mu } |\omega_\mu \rangle_{in} |0\rangle_u + |0\rangle_{in} |\omega_\alpha\rangle_u \right) \\
=&\frac{1}{2}\left[ \sum_\mu C_\mu e^{i \phi_\mu } \left(\hat{a}_r^\dagger (\omega_\mu) + \hat{a}_t^\dagger (\omega_\mu)\right) |0\rangle_{r} |0\rangle_t + 
\left(\hat{a}_r^\dagger (\omega_\alpha) - \hat{a}_t^\dagger (\omega_\alpha) \right) |0\rangle_{r}|0\rangle_t \right] \\
=&\frac{1}{2}\left[ \sum_{\mu \neq \omega_\alpha} C_\mu e^{i \phi_\mu } \left( |\omega_\mu\rangle_{r} |0\rangle_t  +  |0\rangle_{r} |\omega_\mu\rangle_t \right) + 
\left( C_\mu e^{i \phi_\mu } + 1 \right) |\omega_\alpha\rangle_{r}|0\rangle_t + \left( C_\mu e^{i \phi_\mu } - 1 \right) |0\rangle_{r}|\omega_\alpha\rangle_t   \right] .\\
\end{aligned}
\end{equation}
This process allows us to obtain two equations about the probabilities of measurable outcomes: $P\left( |0\rangle_{r}|\omega_\alpha\rangle_t \right)$ and $P\left( |\omega_\alpha\rangle_{r}|0\rangle_t \right)$.
\begin{equation}
\begin{aligned}
& P\left( |0\rangle_{r}|\omega_\alpha\rangle_t \right)= 1 + C_\alpha^2 - 2C_\alpha cos\phi_\alpha ,\\
& P\left( |\omega_\alpha\rangle_{r}|0\rangle_t \right)= 1 + C_\alpha^2 + 2C_\alpha cos\phi_\alpha .\\
\end{aligned}
\end{equation}
We can introduce a $\pi$ phase shift using a delay device.
\begin{equation}
\begin{aligned}
&\frac{1}{\sqrt{2}}\left( \sum_\mu C_\mu e^{i \phi_\mu } |\omega_\mu \rangle_r |0\rangle_t + i |0\rangle_r |\omega_\alpha\rangle_t \right) \\
=&\frac{1}{2}\left[ \sum_{\omega_\mu \neq \omega_\alpha} C_\mu e^{i \phi_\mu } \left( |\omega_\mu\rangle_{r} |0\rangle_t  +  |0\rangle_{r} |\omega_\mu\rangle_t \right) + 
\left( C_\mu e^{i \phi_\mu } + i \right) |\omega_\alpha\rangle_{r}|0\rangle_t + \left( C_\mu e^{i \phi_\mu } - i \right) |0\rangle_{r}|\omega_\alpha\rangle_t   \right], \\
\end{aligned}
\end{equation}
which yields two more equations:
\begin{equation}
\begin{aligned}
& P\left( |0\rangle_{r}|\omega_\alpha\rangle_t \right)= 1 + C_\alpha^2 - 2C_\alpha sin\phi_\alpha , \\
& P\left( |\omega_\alpha\rangle_{r}|0\rangle_t \right)= 1 + C_\alpha^2 + 2C_\alpha sin\phi_\alpha .\\
\end{aligned}
\end{equation}
Therefore, we can calculate the values of $C_\alpha \phi_\alpha$ at different times.  These values, which encode information about the system Hamiltonian, can be used to calculate the SFF in two-photon case.

For example, considering the two-photon scenario, we have
\begin{equation}
\begin{aligned}
|\omega_\alpha\rangle_{in} |\omega_\beta\rangle_u& = a_{in}^\dagger (\omega_\alpha)a_{u}^\dagger (\omega_\beta) |0\rangle_{in} |0\rangle_u\\
& =\frac{1}{2}\left(a_r^\dagger (\omega_\alpha) + a_t^\dagger (\omega_\alpha) \right)  \left(a_r^\dagger (\omega_\beta) - a_t^\dagger (\omega_\beta) \right)  |0\rangle_{r} |0\rangle_t\\
& =\frac{1}{2}\left( |\omega_\alpha\omega_\beta\rangle_r |0\rangle_t - |0\rangle_r |\omega_\alpha\omega_\beta\rangle_t  - |\omega_\alpha\rangle_r |\omega_\beta\rangle_t  + |\omega_\beta\rangle_r |\omega_\alpha\rangle_t \right) .
\end{aligned}
\end{equation}
After evolving, the initial state $|\omega_\alpha\omega_\beta\rangle_r |0\rangle_t$ becomes: $\sum_{\omega_\mu\omega_\nu} C_{\mu\nu} e^{i \phi_{\mu\nu} } |\omega_\mu \omega_\nu\rangle_{in} |0\rangle_u$. Similarly, the states $|\omega_\alpha\rangle_r |\omega_\beta\rangle_t$ and 
$|\omega_\beta\rangle_r |\omega_\alpha\rangle_t$ transform into: $\sum_{\omega_\mu} C_{\mu} e^{i \phi_\mu } |\omega_\mu\rangle_{in}  |\omega_\beta\rangle_u$ and $\sum_{\omega_\mu} C_{\mu} e^{i \phi_\mu } |\omega_\mu\rangle_{in}  |\omega_\alpha\rangle_u$, respectively. The two-photon state exiting BS1 becomes the input for the subsequent recombination process at BS2:
\begin{equation}
\begin{aligned}
&\frac{1}{2}\left[ \sum_{\mu\nu} C_{\mu\nu} e^{i \phi_{\mu\nu} } |\omega_\mu \omega_\nu\rangle_{in}  |0\rangle_u - |0\rangle_{in}  |\omega_\alpha \omega_\beta \rangle_u -\sum_\mu C_\mu e^{i \phi_\mu } |\omega_\mu \rangle_{in}  |\omega_\beta\rangle_u +\sum_\mu C_\mu e^{i \phi_\mu } |\omega_\mu \rangle_{in}  |\omega_\alpha\rangle_u \right] \\
=&\frac{1}{4} \biggl[ 
\sum_{\mu\nu \neq \alpha\beta} C_{\mu\nu} e^{i \phi_{\mu\nu} } \left(
|\omega_\mu\omega_\nu\rangle_{r} |0\rangle_t  +  
|0\rangle_{r} |\omega_\mu\omega_\nu\rangle_t +
|\omega_\mu\rangle_{r} |\omega_\nu\rangle_t +
|\omega_\nu\rangle_{r} |\omega_\mu\rangle_t  \right) \\
& -\sum_{\mu \neq \alpha} C_{\mu} e^{i \phi_{\mu} } \left( 
|\omega_\mu\omega_\nu\rangle_{r} |0\rangle_t -  
|0\rangle_{r} |\omega_\mu\omega_\nu\rangle_t -
|\omega_\mu\rangle_{r} |\omega_\nu\rangle_t +
|\omega_\nu\rangle_{r} |\omega_\mu\rangle_t  \right)\\  
& +\sum_{\mu \neq \beta} C_{\mu} e^{i \phi_{\mu} } \left( 
|\omega_\mu\omega_\nu\rangle_{r} |0\rangle_t -  
|0\rangle_{r} |\omega_\mu\omega_\nu\rangle_t -
|\omega_\mu\rangle_{r} |\omega_\nu\rangle_t +
|\omega_\nu\rangle_{r} |\omega_\mu\rangle_t  \right)\\ 
& +C_{\alpha\beta} e^{i \phi_{\alpha\beta}} \left( 
|\omega_\alpha\omega_\beta\rangle_{r} |0\rangle_t +  
|0\rangle_{r} |\omega_\alpha\omega_\beta\rangle_t +
|\omega_\alpha\rangle_{r} |\omega_\beta\rangle_t +
|\omega_\beta\rangle_{r} |\omega_\alpha\rangle_t  \right)\\ 
& -  \left( 
|\omega_\alpha\omega_\beta\rangle_{r} |0\rangle_t +  
|0\rangle_{r} |\omega_\alpha\omega_\beta\rangle_t -
|\omega_\alpha\rangle_{r} |\omega_\beta\rangle_t -
|\omega_\beta\rangle_{r} |\omega_\alpha\rangle_t  \right)\\ 
& - C_{\alpha} e^{i \phi_{\alpha}} \left( 
|\omega_\alpha\omega_\beta\rangle_{r} |0\rangle_t -  
|0\rangle_{r} |\omega_\alpha\omega_\beta\rangle_t -
|\omega_\alpha\rangle_{r} |\omega_\beta\rangle_t +
|\omega_\beta\rangle_{r} |\omega_\alpha\rangle_t  \right)\\ 
& + C_{\beta} e^{i \phi_{\beta}} \left( 
|\omega_\alpha\omega_\beta\rangle_{r} |0\rangle_t -  
|0\rangle_{r} |\omega_\alpha\omega_\beta\rangle_t -
|\omega_\alpha\rangle_{r} |\omega_\beta\rangle_t +
|\omega_\beta\rangle_{r} |\omega_\alpha\rangle_t  \right)
 \biggr] .
\end{aligned}
\end{equation}

We obtain four the probabilities of finding the photons in specific final states: $ |\omega_\alpha\omega_\beta\rangle_{r}|0\rangle_t $, $ |0\rangle_{r}|\omega_\alpha\omega_\beta\rangle_t $, $ |\omega_\alpha\rangle_{r}|\omega_\beta\rangle_t $ and $|\omega_\beta\rangle_{r}|\omega_\alpha\rangle_t$. 
\begin{equation}
\begin{aligned}
& P\left( |\omega_\alpha\omega_\beta\rangle_{r}|0\rangle_t \right)=\frac{1}{16} |C_{\alpha\beta} e^{i \phi_{\alpha\beta}} - 1 - C_{\alpha} e^{i \phi_{\alpha}} + C_{\beta} e^{i \phi_{\beta}}|^2 ,\\
& P\left( |0\rangle_{r}|\omega_\alpha\omega_\beta\rangle_t \right)=\frac{1}{16} |C_{\alpha\beta} e^{i \phi_{\alpha\beta}} - 1 + C_{\alpha} e^{i \phi_{\alpha}} - C_{\beta} e^{i \phi_{\beta}}|^2 ,\\
& P\left( |\omega_\alpha\rangle_{r}|\omega_\beta\rangle_t \right)=\frac{1}{16} |C_{\alpha\beta} e^{i \phi_{\alpha\beta}} + 1 + C_{\alpha} e^{i \phi_{\alpha}} - C_{\beta} e^{i \phi_{\beta}}|^2 ,\\
& P\left( |\omega_\beta\rangle_{r}|\omega_\alpha\rangle_t \right)=\frac{1}{16} |C_{\alpha\beta} e^{i \phi_{\alpha\beta}} + 1 - C_{\alpha} e^{i \phi_{\alpha}} + C_{\beta} e^{i \phi_{\beta}}|^2 .\\
\end{aligned}
\end{equation}
These probabilities are related to the complex quantities $C_{\alpha\beta} e^{i \phi_{\alpha\beta}}$, $ C_{\alpha} e^{i \phi_{\alpha}}$, $ C_{\beta} e^{i \phi_{\beta}}$. The last two complex quantities are known. So $C_{\alpha\beta} e^{i \phi_{\alpha\beta}}$ could be solved. By performing measurements with various initial two-photon states and collecting the corresponding propagation probabilities, we can reconstruct the SFF.

In this paragraph, we provide analytical expressions for SFF $K(t)$ for various systems.
According to random matrix theory, when a system exhibits chaotic behavior, the $K(t)$ curve typically displays a characteristic ramp and plateau structure. For example, the $K(t)$ function of the GOE is given by~\cite{joshi_probing_2022}:
\begin{equation}
\begin{aligned}
K(t) = r(t)^2   + \frac{1}{D} 
\begin{cases} 
 2\frac{t}{t_H}-\frac{t}{t_H}{\rm log}(1+2\frac{t}{t_H}) & \text{if} \quad 0<t \leq t_H,\\
2-\frac{t}{t_H}{\rm log}(\frac{2t+t_H}{2t-t_H}) & \text{if} \quad t > t_H, \end{cases}
\end{aligned}
\end{equation}
 and the $K(t)$ of the GUE is given by
\begin{equation}
\begin{aligned}
K(t) = r(t)^2  + \frac{1}{D} 
\begin{cases} 
 \frac{t}{t_H} & \text{if} \quad 0<t \leq t_H,\\
1 & \text{if} \quad t > t_H, \end{cases}
\end{aligned}
\end{equation}
where $r(t) = t_H{\rm J_1}(4Dt/t_H)/2Dt))$ and ${\rm J_1}(t)$ is the Bessel function of the first kind. $t_H$ is the Heisenberg time~\cite{joshi_probing_2022}.

\section{Numerical simulation with waveguides and pulse inject}

The initial state is a two-photon pulse Fock state:
\begin{equation}
\begin{aligned}
|\mathrm{ini}\rangle = \int d\omega_n d\omega_m \xi(\omega_n \omega_m) \hat{c}^{\dagger}(\omega_n) c^{\dagger}(\omega_m) |0\rangle.
\end{aligned}
\end{equation}
The normalization of the wave packet requires: 
\begin{equation}
\begin{aligned}
\int d\omega_n d\omega_m |\xi(\omega_n \omega_m)|^2 = 1.
\end{aligned}
\end{equation}
The total Hamiltonian is defined as:
\begin{equation}
\begin{aligned}
\hat{H} = \hat{H}_0 + \hat{H}_{\mathrm{c1}} + \hat{H}_{\mathrm{c2}}  + \hat{H}_\mathrm{b} + \hat{H}_{\mathrm{int-rc1}} + \hat{H}_{\mathrm{int-rc2}} + \hat{H}_{\mathrm{int-rb}} + \hat{H}_{\mathrm{int-bc1}} + \hat{H}_{\mathrm{int-bc2}}.
\end{aligned}
\end{equation}
Here, the $\hat{H}_0$ is the Hamiltonian of ring resonator system. The $\hat{H_{\mathrm{c1}}}$, $\hat{H_{\mathrm{c2}}}$ and $\hat{H_\mathrm{b}}$ describe two waveguide modes and bath modes, respectively. We have
\begin{equation}
\begin{aligned}
\hat{H_{\mathrm{c1}}} = \int d\omega h\omega \hat{c_1}^{\dagger}(\omega) \hat{c_1}(\omega), \quad, \hat{H_{\mathrm{c2}}} = \int d\omega h\omega \hat{c_2}^{\dagger}(\omega) \hat{c_2}(\omega), \quad \hat{H_\mathrm{b}} = \int d\omega h\omega \hat{b}^{\dagger}(\omega) \hat{b}(\omega),
\end{aligned}
\end{equation}
where $\hat{c}_{1,2}^{\dagger}(\omega)$ $\left(\hat{b}^{\dagger}(\omega)\right)$ and $\hat{c}_{1,2}(\omega)$  $\left(\hat{b}(\omega)\right)$ are the creation and annihilation operators of two waveguide modes (bath modes) with frequency $\omega$. 
The $\hat{H}_{\mathrm{int-rc1}}$ and $\hat{H}_{\mathrm{int-rc2}}$ describe the interaction Hamiltonian between ring resonator and two waveguide modes:
\begin{equation}
\begin{aligned}
&\hat{H}_{\mathrm{int-rc1}} = \sum_{n} \int d\omega g_{\mathrm{rc1}}(\omega,\omega_n)\left[\hat{c_1}^{\dagger}(\omega) \hat{a}_n + h.c.\right],\\
&\hat{H}_{\mathrm{int-rc2}} = \sum_{n} \int d\omega g_{\mathrm{rc2}}(\omega,\omega^{\prime})\left[\hat{c_2}^{\dagger}(\omega) \hat{a}(\omega^{\prime}) + h.c.\right].
\end{aligned}
\end{equation}
The $\hat{H}_{\mathrm{int-rb}}$ describes the interaction Hamiltonian between ring resonator and bath modes:
\begin{equation}
\begin{aligned}
&\hat{H}_{\mathrm{int-rb}} = \sum_{n} \int d\omega g_{\mathrm{rb}}(\omega,\omega_n)\left[\hat{b}^{\dagger}(\omega) \hat{a}_n + h.c.\right].
\end{aligned}
\end{equation}
The $\hat{H}_{\mathrm{int-bc1}}$ and $\hat{H}_{\mathrm{int-bc2}}$ describe the interaction Hamiltonian between bath and two waveguide modes:
\begin{equation}
\begin{aligned}
&\hat{H}_{\mathrm{int-bc1}} = \sum_{n} \int d\omega g_{\mathrm{bc1}}(\omega,\omega^{\prime})\left[\hat{c_1}^{\dagger}(\omega) \hat{b}(\omega^{\prime}) + h.c.\right],\\ 
&\hat{H}_{\mathrm{int-bc2}} = \sum_{n} \int d\omega g_{\mathrm{bc2}}(\omega,\omega^{\prime})\left[\hat{c_2}^{\dagger}(\omega) \hat{b}(\omega^{\prime}) + h.c.\right].\\ 
\end{aligned}
\end{equation}
Here, $g_{\mathrm{rc1}}(\omega,\omega_n)$, $g_{\mathrm{bc1}}(\omega,\omega^{\prime})$, $g_{\mathrm{rc2}}(\omega,\omega_n)$, $g_{\mathrm{bc2}}(\omega,\omega^{\prime})$, $g_{\mathrm{bc2}}(\omega,\omega_n)$ and $g_{\mathrm{rb}}(\omega,\omega_n)$ are the coupling amplitudes between ring resonator modes and input waveguide modes, input waveguide modes and bath, ring resonator modes and output waveguide modes, output waveguide modes and bath as well as ring resonator modes and bath. They having Lorentzian form~\cite{PhysRevA.Markovian}:
\begin{equation}
\begin{aligned}
&|g_{\mathrm{rc1}}(\omega,\omega_n)|^2 = \frac{\gamma_{\mathrm{rc1}}/2\pi}{[(\omega-\omega_n)/\kappa_{\mathrm{rc1}}]^2+1},\\
&|g_{\mathrm{bc1}}(\omega,\omega^{\prime})|^2 = \frac{\gamma_{\mathrm{bc1}}/2\pi}{[(\omega-\omega_n)/\kappa_{\mathrm{bc1}}]^2+1},\\
&|g_{\mathrm{rc2}}(\omega,\omega_n)|^2 = \frac{\gamma_{\mathrm{rc2}}/2\pi}{[(\omega-\omega_n)/\kappa_{\mathrm{rc2}}]^2+1},\\
&|g_{\mathrm{bc2}}(\omega,\omega_n)|^2 = \frac{\gamma_{\mathrm{bc2}}/2\pi}{[(\omega-\omega_n)/\kappa_{\mathrm{bc2}}]^2+1},\\
&|g_{\mathrm{rb}}(\omega,\omega_n)|^2 = \frac{\gamma_{\mathrm{rb}}/2\pi}{[(\omega-\omega_n)/\kappa_{\mathrm{rb}}]^2+1},
\end{aligned}
\end{equation}
where $\gamma_{\mathrm{rc1}}$, $\gamma_{\mathrm{bc1}}$, $\gamma_{\mathrm{rc2}}$, $\gamma_{\mathrm{bc2}}$, $\gamma_{\mathrm{rb}}$ characterize the decay of the ring resonator modes to input waveguide modes, input waveguide modes to bath, ring resonator modes to output waveguide modes, output waveguide modes to bath and ring resonator modes to bath. The $\kappa_{\mathrm{rc1}}$, $\kappa_{\mathrm{bc1}}$, $\kappa_{\mathrm{rc2}}$, $\kappa_{\mathrm{bc2}}$, $\kappa_{\mathrm{rb}}$ are determined by the Q-factors of the ring resonator, input waveguide and output waveguide.

In the regime of $\gamma_{\mathrm{bc1}}, \gamma_{\mathrm{bc2}} \gg  \gamma_{\mathrm{rc1}}, \gamma_{\mathrm{rc2}}$, we can make an approximation that the system evolution can be separated into two processes: photon injection into the ring resonator and subsequent photon evolution within the ring. For the former one, due to the high dissipation coefficient in the waveguide, the ring resonant interacts with a double-photon wave packet that decays exponentially with time: $|\psi_\mathrm{in}\rangle = \int d\omega_n d\omega_m e^{-\gamma_{\mathrm{bc1}} t} \xi(\omega_n \omega_m) \hat{c}^{\dagger}(\omega_n) c^{\dagger}(\omega_m) |0\rangle$. Due to the large value of $\gamma_{\mathrm{bc1}}$ compared to $g$, the injection of photons into the ring resonator happens on a much faster timescale than the overall evolution of the system. This rapid injection process effectively excites the ring resonator modes with the two-photon Fock state initially present in the waveguide modes. After this process, the ring resonant obtains a two-photon wave packet with slightly broadened distribution. However, due to the dissipation, the expected number of photons in the photon state is less than $2$. For the latter one, we ignore the waveguide modes and regards that as dissipation of ring resonator modes. Then, we have the master equation to describe the evolution of the ring resonator modes as
\begin{equation}
\begin{aligned}
\dot\rho_0(t) = {-i}[\hat{H}_0,\rho_0] + \Gamma \sum_n \left( a_n \rho_0 a^{\dagger}_n - \frac{1}{2} a^{\dagger}_n a_n \rho_0 - \frac{1}{2} \rho_0 a^{\dagger}_n a_n \right),
\end{aligned}
\end{equation}
where $\Gamma = \gamma_{\mathrm{rc1}} + \gamma_{\mathrm{rc2}} + \gamma_{\mathrm{rb}}$ is the total loss efficiency~\cite{Breuer2007, PhysRevA.79.023838, PhysRevA.82.063821}.

\end{onecolumngrid}


\begin{thebibliography}{63}%
\makeatletter
\providecommand \@ifxundefined [1]{%
 \@ifx{#1\undefined}
}%
\providecommand \@ifnum [1]{%
 \ifnum #1\expandafter \@firstoftwo
 \else \expandafter \@secondoftwo
 \fi
}%
\providecommand \@ifx [1]{%
 \ifx #1\expandafter \@firstoftwo
 \else \expandafter \@secondoftwo
 \fi
}%
\providecommand \natexlab [1]{#1}%
\providecommand \enquote  [1]{``#1''}%
\providecommand \bibnamefont  [1]{#1}%
\providecommand \bibfnamefont [1]{#1}%
\providecommand \citenamefont [1]{#1}%
\providecommand \href@noop [0]{\@secondoftwo}%
\providecommand \href [0]{\begingroup \@sanitize@url \@href}%
\providecommand \@href[1]{\@@startlink{#1}\@@href}%
\providecommand \@@href[1]{\endgroup#1\@@endlink}%
\providecommand \@sanitize@url [0]{\catcode `\\12\catcode `\$12\catcode
  `\&12\catcode `\#12\catcode `\^12\catcode `\_12\catcode `\%12\relax}%
\providecommand \@@startlink[1]{}%
\providecommand \@@endlink[0]{}%
\providecommand \url  [0]{\begingroup\@sanitize@url \@url }%
\providecommand \@url [1]{\endgroup\@href {#1}{\urlprefix }}%
\providecommand \urlprefix  [0]{URL }%
\providecommand \Eprint [0]{\href }%
\providecommand \doibase [0]{https://doi.org/}%
\providecommand \selectlanguage [0]{\@gobble}%
\providecommand \bibinfo  [0]{\@secondoftwo}%
\providecommand \bibfield  [0]{\@secondoftwo}%
\providecommand \translation [1]{[#1]}%
\providecommand \BibitemOpen [0]{}%
\providecommand \bibitemStop [0]{}%
\providecommand \bibitemNoStop [0]{.\EOS\space}%
\providecommand \EOS [0]{\spacefactor3000\relax}%
\providecommand \BibitemShut  [1]{\csname bibitem#1\endcsname}%
\let\auto@bib@innerbib\@empty
\bibitem [{\citenamefont {Jozsa}\ and\ \citenamefont
  {Linden}(2003)}]{jozsa2003role}%
  \BibitemOpen
  \bibfield  {author} {\bibinfo {author} {\bibfnamefont {R.}~\bibnamefont
  {Jozsa}}\ and\ \bibinfo {author} {\bibfnamefont {N.}~\bibnamefont {Linden}},\
  }\href {https://doi.org/10.1098/rspa.2002.1097} {\bibfield  {journal}
  {\bibinfo  {journal} {Proceedings of the Royal Society of London. Series A:
  Mathematical, Physical and Engineering Sciences}\ }\textbf {\bibinfo {volume}
  {459}},\ \bibinfo {pages} {2011} (\bibinfo {year} {2003})}\BibitemShut
  {NoStop}%
\bibitem [{\citenamefont {Bouwmeester}\ \emph {et~al.}(1997)\citenamefont
  {Bouwmeester}, \citenamefont {Pan}, \citenamefont {Mattle}, \citenamefont
  {Eibl}, \citenamefont {Weinfurter},\ and\ \citenamefont
  {Zeilinger}}]{Bouwmeester1997}%
  \BibitemOpen
  \bibfield  {author} {\bibinfo {author} {\bibfnamefont {D.}~\bibnamefont
  {Bouwmeester}}, \bibinfo {author} {\bibfnamefont {J.-W.}\ \bibnamefont
  {Pan}}, \bibinfo {author} {\bibfnamefont {K.}~\bibnamefont {Mattle}},
  \bibinfo {author} {\bibfnamefont {M.}~\bibnamefont {Eibl}}, \bibinfo {author}
  {\bibfnamefont {H.}~\bibnamefont {Weinfurter}},\ and\ \bibinfo {author}
  {\bibfnamefont {A.}~\bibnamefont {Zeilinger}},\ }\href
  {https://doi.org/10.1038/37539} {\bibfield  {journal} {\bibinfo  {journal}
  {Nature}\ }\textbf {\bibinfo {volume} {390}},\ \bibinfo {pages} {575}
  (\bibinfo {year} {1997})}\BibitemShut {NoStop}%
\bibitem [{\citenamefont {Gao}\ \emph {et~al.}(2005)\citenamefont {Gao},
  \citenamefont {Yan},\ and\ \citenamefont {Wang}}]{gao2005deterministic}%
  \BibitemOpen
  \bibfield  {author} {\bibinfo {author} {\bibfnamefont {T.}~\bibnamefont
  {Gao}}, \bibinfo {author} {\bibfnamefont {F.~L.}\ \bibnamefont {Yan}},\ and\
  \bibinfo {author} {\bibfnamefont {Z.~X.}\ \bibnamefont {Wang}},\ }\href
  {https://doi.org/10.1088/0305-4470/38/25/011} {\bibfield  {journal} {\bibinfo
   {journal} {Journal of Physics A: Mathematical and General}\ }\textbf
  {\bibinfo {volume} {38}},\ \bibinfo {pages} {5761} (\bibinfo {year}
  {2005})}\BibitemShut {NoStop}%
\bibitem [{\citenamefont {Su}\ \emph {et~al.}(2016)\citenamefont {Su},
  \citenamefont {Tian}, \citenamefont {Deng}, \citenamefont {Li}, \citenamefont
  {Xie},\ and\ \citenamefont {Peng}}]{su2016quantum}%
  \BibitemOpen
  \bibfield  {author} {\bibinfo {author} {\bibfnamefont {X.}~\bibnamefont
  {Su}}, \bibinfo {author} {\bibfnamefont {C.}~\bibnamefont {Tian}}, \bibinfo
  {author} {\bibfnamefont {X.}~\bibnamefont {Deng}}, \bibinfo {author}
  {\bibfnamefont {Q.}~\bibnamefont {Li}}, \bibinfo {author} {\bibfnamefont
  {C.}~\bibnamefont {Xie}},\ and\ \bibinfo {author} {\bibfnamefont
  {K.}~\bibnamefont {Peng}},\ }\href
  {https://doi.org/10.1103/PhysRevLett.117.240503} {\bibfield  {journal}
  {\bibinfo  {journal} {Phys. Rev. Lett.}\ }\textbf {\bibinfo {volume} {117}},\
  \bibinfo {pages} {240503} (\bibinfo {year} {2016})}\BibitemShut {NoStop}%
\bibitem [{\citenamefont {Shannon}\ \emph {et~al.}(2020)\citenamefont
  {Shannon}, \citenamefont {Towe},\ and\ \citenamefont
  {Tonguz}}]{shannon2020use}%
  \BibitemOpen
  \bibfield  {author} {\bibinfo {author} {\bibfnamefont {K.}~\bibnamefont
  {Shannon}}, \bibinfo {author} {\bibfnamefont {E.}~\bibnamefont {Towe}},\ and\
  \bibinfo {author} {\bibfnamefont {O.~K.}\ \bibnamefont {Tonguz}},\ }\href
  {https://api.semanticscholar.org/CorpusID:212747977} {\bibfield  {journal}
  {\bibinfo  {journal} {ArXiv}\ }\textbf {\bibinfo {volume} {abs/2003.07907}}
  (\bibinfo {year} {2020})}\BibitemShut {NoStop}%
\bibitem [{\citenamefont {Long}\ \emph {et~al.}(2007)\citenamefont {Long},
  \citenamefont {Deng}, \citenamefont {Wang}, \citenamefont {Li}, \citenamefont
  {Wen},\ and\ \citenamefont {Wang}}]{long2007quantum}%
  \BibitemOpen
  \bibfield  {author} {\bibinfo {author} {\bibfnamefont {G.-l.}\ \bibnamefont
  {Long}}, \bibinfo {author} {\bibfnamefont {F.-g.}\ \bibnamefont {Deng}},
  \bibinfo {author} {\bibfnamefont {C.}~\bibnamefont {Wang}}, \bibinfo {author}
  {\bibfnamefont {X.-h.}\ \bibnamefont {Li}}, \bibinfo {author} {\bibfnamefont
  {K.}~\bibnamefont {Wen}},\ and\ \bibinfo {author} {\bibfnamefont {W.-y.}\
  \bibnamefont {Wang}},\ }\href {https://doi.org/10.1007/s11467-007-0050-3}
  {\bibfield  {journal} {\bibinfo  {journal} {Frontiers of Physics in China}\
  }\textbf {\bibinfo {volume} {2}},\ \bibinfo {pages} {251} (\bibinfo {year}
  {2007})}\BibitemShut {NoStop}%
\bibitem [{\citenamefont {Qi}\ \emph {et~al.}(2019)\citenamefont {Qi},
  \citenamefont {Sun}, \citenamefont {Lin}, \citenamefont {Niu}, \citenamefont
  {Hao}, \citenamefont {Song}, \citenamefont {Huang}, \citenamefont {Gao},
  \citenamefont {Yin},\ and\ \citenamefont {Long}}]{qi2019implementation}%
  \BibitemOpen
  \bibfield  {author} {\bibinfo {author} {\bibfnamefont {R.}~\bibnamefont
  {Qi}}, \bibinfo {author} {\bibfnamefont {Z.}~\bibnamefont {Sun}}, \bibinfo
  {author} {\bibfnamefont {Z.}~\bibnamefont {Lin}}, \bibinfo {author}
  {\bibfnamefont {P.}~\bibnamefont {Niu}}, \bibinfo {author} {\bibfnamefont
  {W.}~\bibnamefont {Hao}}, \bibinfo {author} {\bibfnamefont {L.}~\bibnamefont
  {Song}}, \bibinfo {author} {\bibfnamefont {Q.}~\bibnamefont {Huang}},
  \bibinfo {author} {\bibfnamefont {J.}~\bibnamefont {Gao}}, \bibinfo {author}
  {\bibfnamefont {L.}~\bibnamefont {Yin}},\ and\ \bibinfo {author}
  {\bibfnamefont {G.-L.}\ \bibnamefont {Long}},\ }\href
  {https://doi.org/10.1038/s41377-019-0132-3} {\bibfield  {journal} {\bibinfo
  {journal} {Light: Science {\&} Applications}\ }\textbf {\bibinfo {volume}
  {8}},\ \bibinfo {pages} {22} (\bibinfo {year} {2019})}\BibitemShut {NoStop}%
\bibitem [{\citenamefont {Portmann}\ and\ \citenamefont
  {Renner}(2022)}]{portmann2022security}%
  \BibitemOpen
  \bibfield  {author} {\bibinfo {author} {\bibfnamefont {C.}~\bibnamefont
  {Portmann}}\ and\ \bibinfo {author} {\bibfnamefont {R.}~\bibnamefont
  {Renner}},\ }\href {https://doi.org/10.1103/RevModPhys.94.025008} {\bibfield
  {journal} {\bibinfo  {journal} {Rev. Mod. Phys.}\ }\textbf {\bibinfo {volume}
  {94}},\ \bibinfo {pages} {025008} (\bibinfo {year} {2022})}\BibitemShut
  {NoStop}%
\bibitem [{\citenamefont {Para{\"i}so}\ \emph {et~al.}(2021)\citenamefont
  {Para{\"i}so}, \citenamefont {Roger}, \citenamefont {Marangon}, \citenamefont
  {De~Marco}, \citenamefont {Sanzaro}, \citenamefont {Woodward}, \citenamefont
  {Dynes}, \citenamefont {Yuan},\ and\ \citenamefont
  {Shields}}]{paraiso2021photonic}%
  \BibitemOpen
  \bibfield  {author} {\bibinfo {author} {\bibfnamefont {T.~K.}\ \bibnamefont
  {Para{\"i}so}}, \bibinfo {author} {\bibfnamefont {T.}~\bibnamefont {Roger}},
  \bibinfo {author} {\bibfnamefont {D.~G.}\ \bibnamefont {Marangon}}, \bibinfo
  {author} {\bibfnamefont {I.}~\bibnamefont {De~Marco}}, \bibinfo {author}
  {\bibfnamefont {M.}~\bibnamefont {Sanzaro}}, \bibinfo {author} {\bibfnamefont
  {R.~I.}\ \bibnamefont {Woodward}}, \bibinfo {author} {\bibfnamefont {J.~F.}\
  \bibnamefont {Dynes}}, \bibinfo {author} {\bibfnamefont {Z.}~\bibnamefont
  {Yuan}},\ and\ \bibinfo {author} {\bibfnamefont {A.~J.}\ \bibnamefont
  {Shields}},\ }\href {https://doi.org/10.1038/s41566-021-00873-0} {\bibfield
  {journal} {\bibinfo  {journal} {Nature Photonics}\ }\textbf {\bibinfo
  {volume} {15}},\ \bibinfo {pages} {850} (\bibinfo {year} {2021})}\BibitemShut
  {NoStop}%
\bibitem [{\citenamefont {Mair}\ \emph {et~al.}(2001)\citenamefont {Mair},
  \citenamefont {Vaziri}, \citenamefont {Weihs},\ and\ \citenamefont
  {Zeilinger}}]{Mair2001}%
  \BibitemOpen
  \bibfield  {author} {\bibinfo {author} {\bibfnamefont {A.}~\bibnamefont
  {Mair}}, \bibinfo {author} {\bibfnamefont {A.}~\bibnamefont {Vaziri}},
  \bibinfo {author} {\bibfnamefont {G.}~\bibnamefont {Weihs}},\ and\ \bibinfo
  {author} {\bibfnamefont {A.}~\bibnamefont {Zeilinger}},\ }\href
  {https://doi.org/10.1038/35085529} {\bibfield  {journal} {\bibinfo  {journal}
  {Nature}\ }\textbf {\bibinfo {volume} {412}},\ \bibinfo {pages} {313}
  (\bibinfo {year} {2001})}\BibitemShut {NoStop}%
\bibitem [{\citenamefont {Defienne}\ \emph {et~al.}(2021)\citenamefont
  {Defienne}, \citenamefont {Ndagano}, \citenamefont {Lyons},\ and\
  \citenamefont {Faccio}}]{Defienne2021}%
  \BibitemOpen
  \bibfield  {author} {\bibinfo {author} {\bibfnamefont {H.}~\bibnamefont
  {Defienne}}, \bibinfo {author} {\bibfnamefont {B.}~\bibnamefont {Ndagano}},
  \bibinfo {author} {\bibfnamefont {A.}~\bibnamefont {Lyons}},\ and\ \bibinfo
  {author} {\bibfnamefont {D.}~\bibnamefont {Faccio}},\ }\href
  {https://doi.org/10.1038/s41567-020-01156-1} {\bibfield  {journal} {\bibinfo
  {journal} {Nature Physics}\ }\textbf {\bibinfo {volume} {17}},\ \bibinfo
  {pages} {591} (\bibinfo {year} {2021})}\BibitemShut {NoStop}%
\bibitem [{\citenamefont {Cheng}\ \emph
  {et~al.}(2023{\natexlab{a}})\citenamefont {Cheng}, \citenamefont {Chang},
  \citenamefont {Sarihan}, \citenamefont {Mueller}, \citenamefont {Spiropulu},
  \citenamefont {Shaw}, \citenamefont {Korzh}, \citenamefont {Faraon},
  \citenamefont {Wong}, \citenamefont {Shapiro},\ and\ \citenamefont
  {Wong}}]{Cheng2023}%
  \BibitemOpen
  \bibfield  {author} {\bibinfo {author} {\bibfnamefont {X.}~\bibnamefont
  {Cheng}}, \bibinfo {author} {\bibfnamefont {K.-C.}\ \bibnamefont {Chang}},
  \bibinfo {author} {\bibfnamefont {M.~C.}\ \bibnamefont {Sarihan}}, \bibinfo
  {author} {\bibfnamefont {A.}~\bibnamefont {Mueller}}, \bibinfo {author}
  {\bibfnamefont {M.}~\bibnamefont {Spiropulu}}, \bibinfo {author}
  {\bibfnamefont {M.~D.}\ \bibnamefont {Shaw}}, \bibinfo {author}
  {\bibfnamefont {B.}~\bibnamefont {Korzh}}, \bibinfo {author} {\bibfnamefont
  {A.}~\bibnamefont {Faraon}}, \bibinfo {author} {\bibfnamefont {F.~N.~C.}\
  \bibnamefont {Wong}}, \bibinfo {author} {\bibfnamefont {J.~H.}\ \bibnamefont
  {Shapiro}},\ and\ \bibinfo {author} {\bibfnamefont {C.~W.}\ \bibnamefont
  {Wong}},\ }\href {https://doi.org/10.1038/s42005-023-01370-2} {\bibfield
  {journal} {\bibinfo  {journal} {Communications Physics}\ }\textbf {\bibinfo
  {volume} {6}},\ \bibinfo {pages} {278} (\bibinfo {year}
  {2023}{\natexlab{a}})}\BibitemShut {NoStop}%
\bibitem [{\citenamefont {Yuan}\ \emph {et~al.}(2016)\citenamefont {Yuan},
  \citenamefont {Shi},\ and\ \citenamefont {Fan}}]{Yuan:16}%
  \BibitemOpen
  \bibfield  {author} {\bibinfo {author} {\bibfnamefont {L.}~\bibnamefont
  {Yuan}}, \bibinfo {author} {\bibfnamefont {Y.}~\bibnamefont {Shi}},\ and\
  \bibinfo {author} {\bibfnamefont {S.}~\bibnamefont {Fan}},\ }\href
  {https://doi.org/10.1364/OL.41.000741} {\bibfield  {journal} {\bibinfo
  {journal} {Opt. Lett.}\ }\textbf {\bibinfo {volume} {41}},\ \bibinfo {pages}
  {741} (\bibinfo {year} {2016})}\BibitemShut {NoStop}%
\bibitem [{\citenamefont {Ozawa}\ \emph {et~al.}(2016)\citenamefont {Ozawa},
  \citenamefont {Price}, \citenamefont {Goldman}, \citenamefont {Zilberberg},\
  and\ \citenamefont {Carusotto}}]{PhysRevA.93.043827}%
  \BibitemOpen
  \bibfield  {author} {\bibinfo {author} {\bibfnamefont {T.}~\bibnamefont
  {Ozawa}}, \bibinfo {author} {\bibfnamefont {H.~M.}\ \bibnamefont {Price}},
  \bibinfo {author} {\bibfnamefont {N.}~\bibnamefont {Goldman}}, \bibinfo
  {author} {\bibfnamefont {O.}~\bibnamefont {Zilberberg}},\ and\ \bibinfo
  {author} {\bibfnamefont {I.}~\bibnamefont {Carusotto}},\ }\href
  {https://doi.org/10.1103/PhysRevA.93.043827} {\bibfield  {journal} {\bibinfo
  {journal} {Phys. Rev. A}\ }\textbf {\bibinfo {volume} {93}},\ \bibinfo
  {pages} {043827} (\bibinfo {year} {2016})}\BibitemShut {NoStop}%
\bibitem [{\citenamefont {Lustig}\ \emph
  {et~al.}(2019{\natexlab{a}})\citenamefont {Lustig}, \citenamefont {Weimann},
  \citenamefont {Plotnik}, \citenamefont {Lumer}, \citenamefont {Bandres},
  \citenamefont {Szameit},\ and\ \citenamefont {Segev}}]{Lustig2019}%
  \BibitemOpen
  \bibfield  {author} {\bibinfo {author} {\bibfnamefont {E.}~\bibnamefont
  {Lustig}}, \bibinfo {author} {\bibfnamefont {S.}~\bibnamefont {Weimann}},
  \bibinfo {author} {\bibfnamefont {Y.}~\bibnamefont {Plotnik}}, \bibinfo
  {author} {\bibfnamefont {Y.}~\bibnamefont {Lumer}}, \bibinfo {author}
  {\bibfnamefont {M.~A.}\ \bibnamefont {Bandres}}, \bibinfo {author}
  {\bibfnamefont {A.}~\bibnamefont {Szameit}},\ and\ \bibinfo {author}
  {\bibfnamefont {M.}~\bibnamefont {Segev}},\ }\href
  {https://doi.org/10.1038/s41586-019-0943-7} {\bibfield  {journal} {\bibinfo
  {journal} {Nature}\ }\textbf {\bibinfo {volume} {567}},\ \bibinfo {pages}
  {356} (\bibinfo {year} {2019}{\natexlab{a}})}\BibitemShut {NoStop}%
\bibitem [{\citenamefont {Luo}\ \emph {et~al.}(2015)\citenamefont {Luo},
  \citenamefont {Zhou}, \citenamefont {Li}, \citenamefont {Xu}, \citenamefont
  {Guo},\ and\ \citenamefont {Zhou}}]{Luo2015}%
  \BibitemOpen
  \bibfield  {author} {\bibinfo {author} {\bibfnamefont {X.-W.}\ \bibnamefont
  {Luo}}, \bibinfo {author} {\bibfnamefont {X.}~\bibnamefont {Zhou}}, \bibinfo
  {author} {\bibfnamefont {C.-F.}\ \bibnamefont {Li}}, \bibinfo {author}
  {\bibfnamefont {J.-S.}\ \bibnamefont {Xu}}, \bibinfo {author} {\bibfnamefont
  {G.-C.}\ \bibnamefont {Guo}},\ and\ \bibinfo {author} {\bibfnamefont {Z.-W.}\
  \bibnamefont {Zhou}},\ }\href {https://doi.org/10.1038/ncomms8704} {\bibfield
   {journal} {\bibinfo  {journal} {Nature Communications}\ }\textbf {\bibinfo
  {volume} {6}},\ \bibinfo {pages} {7704} (\bibinfo {year} {2015})}\BibitemShut
  {NoStop}%
\bibitem [{\citenamefont {Javid}\ \emph {et~al.}(2023)\citenamefont {Javid},
  \citenamefont {Lopez-Rios}, \citenamefont {Ling}, \citenamefont {Graf},
  \citenamefont {Staffa},\ and\ \citenamefont {Lin}}]{Javid2023}%
  \BibitemOpen
  \bibfield  {author} {\bibinfo {author} {\bibfnamefont {U.~A.}\ \bibnamefont
  {Javid}}, \bibinfo {author} {\bibfnamefont {R.}~\bibnamefont {Lopez-Rios}},
  \bibinfo {author} {\bibfnamefont {J.}~\bibnamefont {Ling}}, \bibinfo {author}
  {\bibfnamefont {A.}~\bibnamefont {Graf}}, \bibinfo {author} {\bibfnamefont
  {J.}~\bibnamefont {Staffa}},\ and\ \bibinfo {author} {\bibfnamefont
  {Q.}~\bibnamefont {Lin}},\ }\href
  {https://doi.org/10.1038/s41566-023-01236-7} {\bibfield  {journal} {\bibinfo
  {journal} {Nature Photonics}\ }\textbf {\bibinfo {volume} {17}},\ \bibinfo
  {pages} {883} (\bibinfo {year} {2023})}\BibitemShut {NoStop}%
\bibitem [{\citenamefont {Xiao}\ \emph {et~al.}(2022)\citenamefont {Xiao},
  \citenamefont {Wang}, \citenamefont {Li}, \citenamefont {Chen},\ and\
  \citenamefont {Yuan}}]{Xiao2022}%
  \BibitemOpen
  \bibfield  {author} {\bibinfo {author} {\bibfnamefont {H.}~\bibnamefont
  {Xiao}}, \bibinfo {author} {\bibfnamefont {L.}~\bibnamefont {Wang}}, \bibinfo
  {author} {\bibfnamefont {Z.-H.}\ \bibnamefont {Li}}, \bibinfo {author}
  {\bibfnamefont {X.}~\bibnamefont {Chen}},\ and\ \bibinfo {author}
  {\bibfnamefont {L.}~\bibnamefont {Yuan}},\ }\href
  {https://doi.org/10.1038/s41534-022-00591-7} {\bibfield  {journal} {\bibinfo
  {journal} {npj Quantum Information}\ }\textbf {\bibinfo {volume} {8}},\
  \bibinfo {pages} {80} (\bibinfo {year} {2022})}\BibitemShut {NoStop}%
\bibitem [{\citenamefont {Du}\ \emph {et~al.}(2022)\citenamefont {Du},
  \citenamefont {Zhang}, \citenamefont {Wu}, \citenamefont {Kockum},\ and\
  \citenamefont {Li}}]{PhysRevLett.128.223602}%
  \BibitemOpen
  \bibfield  {author} {\bibinfo {author} {\bibfnamefont {L.}~\bibnamefont
  {Du}}, \bibinfo {author} {\bibfnamefont {Y.}~\bibnamefont {Zhang}}, \bibinfo
  {author} {\bibfnamefont {J.-H.}\ \bibnamefont {Wu}}, \bibinfo {author}
  {\bibfnamefont {A.~F.}\ \bibnamefont {Kockum}},\ and\ \bibinfo {author}
  {\bibfnamefont {Y.}~\bibnamefont {Li}},\ }\href
  {https://doi.org/10.1103/PhysRevLett.128.223602} {\bibfield  {journal}
  {\bibinfo  {journal} {Phys. Rev. Lett.}\ }\textbf {\bibinfo {volume} {128}},\
  \bibinfo {pages} {223602} (\bibinfo {year} {2022})}\BibitemShut {NoStop}%
\bibitem [{\citenamefont {Yuan}\ \emph {et~al.}(2020)\citenamefont {Yuan},
  \citenamefont {Dutt}, \citenamefont {Qin}, \citenamefont {Fan},\ and\
  \citenamefont {Chen}}]{yuan_creating_2020}%
  \BibitemOpen
  \bibfield  {author} {\bibinfo {author} {\bibfnamefont {L.}~\bibnamefont
  {Yuan}}, \bibinfo {author} {\bibfnamefont {A.}~\bibnamefont {Dutt}}, \bibinfo
  {author} {\bibfnamefont {M.}~\bibnamefont {Qin}}, \bibinfo {author}
  {\bibfnamefont {S.}~\bibnamefont {Fan}},\ and\ \bibinfo {author}
  {\bibfnamefont {X.}~\bibnamefont {Chen}},\ }\href
  {https://doi.org/10.1364/PRJ.396731} {\bibfield  {journal} {\bibinfo
  {journal} {Photon. Res.}\ }\textbf {\bibinfo {volume} {8}},\ \bibinfo {pages}
  {B8} (\bibinfo {year} {2020})},\ \bibinfo {note} {publisher: Optica
  Publishing Group}\BibitemShut {NoStop}%
\bibitem [{\citenamefont {Tusnin}\ \emph {et~al.}(2020)\citenamefont {Tusnin},
  \citenamefont {Tikan},\ and\ \citenamefont
  {Kippenberg}}]{PhysRevA.102.023518}%
  \BibitemOpen
  \bibfield  {author} {\bibinfo {author} {\bibfnamefont {A.~K.}\ \bibnamefont
  {Tusnin}}, \bibinfo {author} {\bibfnamefont {A.~M.}\ \bibnamefont {Tikan}},\
  and\ \bibinfo {author} {\bibfnamefont {T.~J.}\ \bibnamefont {Kippenberg}},\
  }\href {https://doi.org/10.1103/PhysRevA.102.023518} {\bibfield  {journal}
  {\bibinfo  {journal} {Phys. Rev. A}\ }\textbf {\bibinfo {volume} {102}},\
  \bibinfo {pages} {023518} (\bibinfo {year} {2020})}\BibitemShut {NoStop}%
\bibitem [{\citenamefont {Englebert}\ \emph {et~al.}(2023)\citenamefont
  {Englebert}, \citenamefont {Goldman}, \citenamefont {Erkintalo},
  \citenamefont {Mostaan}, \citenamefont {Gorza}, \citenamefont {Leo},\ and\
  \citenamefont {Fatome}}]{Englebert2023}%
  \BibitemOpen
  \bibfield  {author} {\bibinfo {author} {\bibfnamefont {N.}~\bibnamefont
  {Englebert}}, \bibinfo {author} {\bibfnamefont {N.}~\bibnamefont {Goldman}},
  \bibinfo {author} {\bibfnamefont {M.}~\bibnamefont {Erkintalo}}, \bibinfo
  {author} {\bibfnamefont {N.}~\bibnamefont {Mostaan}}, \bibinfo {author}
  {\bibfnamefont {S.-P.}\ \bibnamefont {Gorza}}, \bibinfo {author}
  {\bibfnamefont {F.}~\bibnamefont {Leo}},\ and\ \bibinfo {author}
  {\bibfnamefont {J.}~\bibnamefont {Fatome}},\ }\href
  {https://doi.org/10.1038/s41567-023-02005-7} {\bibfield  {journal} {\bibinfo
  {journal} {Nature Physics}\ }\textbf {\bibinfo {volume} {19}},\ \bibinfo
  {pages} {1014} (\bibinfo {year} {2023})}\BibitemShut {NoStop}%
\bibitem [{\citenamefont {Yuan}\ \emph {et~al.}(2018)\citenamefont {Yuan},
  \citenamefont {Lin}, \citenamefont {Xiao},\ and\ \citenamefont
  {Fan}}]{Yuan:18}%
  \BibitemOpen
  \bibfield  {author} {\bibinfo {author} {\bibfnamefont {L.}~\bibnamefont
  {Yuan}}, \bibinfo {author} {\bibfnamefont {Q.}~\bibnamefont {Lin}}, \bibinfo
  {author} {\bibfnamefont {M.}~\bibnamefont {Xiao}},\ and\ \bibinfo {author}
  {\bibfnamefont {S.}~\bibnamefont {Fan}},\ }\href
  {https://doi.org/10.1364/OPTICA.5.001396} {\bibfield  {journal} {\bibinfo
  {journal} {Optica}\ }\textbf {\bibinfo {volume} {5}},\ \bibinfo {pages}
  {1396} (\bibinfo {year} {2018})}\BibitemShut {NoStop}%
\bibitem [{\citenamefont {Yuan}\ \emph {et~al.}(2021)\citenamefont {Yuan},
  \citenamefont {Dutt},\ and\ \citenamefont {Fan}}]{Yuan2021}%
  \BibitemOpen
  \bibfield  {author} {\bibinfo {author} {\bibfnamefont {L.}~\bibnamefont
  {Yuan}}, \bibinfo {author} {\bibfnamefont {A.}~\bibnamefont {Dutt}},\ and\
  \bibinfo {author} {\bibfnamefont {S.}~\bibnamefont {Fan}},\ }\href
  {https://doi.org/10.1063/5.0056359} {\bibfield  {journal} {\bibinfo
  {journal} {APL Photonics}\ }\textbf {\bibinfo {volume} {6}},\ \bibinfo
  {pages} {071102} (\bibinfo {year} {2021})}\BibitemShut {NoStop}%
\bibitem [{\citenamefont {Heckelmann}\ \emph {et~al.}(2023)\citenamefont
  {Heckelmann}, \citenamefont {Bertrand}, \citenamefont {Dikopoltsev},
  \citenamefont {Beck}, \citenamefont {Scalari},\ and\ \citenamefont
  {Faist}}]{Heckelmann2023}%
  \BibitemOpen
  \bibfield  {author} {\bibinfo {author} {\bibfnamefont {I.}~\bibnamefont
  {Heckelmann}}, \bibinfo {author} {\bibfnamefont {M.}~\bibnamefont
  {Bertrand}}, \bibinfo {author} {\bibfnamefont {A.}~\bibnamefont
  {Dikopoltsev}}, \bibinfo {author} {\bibfnamefont {M.}~\bibnamefont {Beck}},
  \bibinfo {author} {\bibfnamefont {G.}~\bibnamefont {Scalari}},\ and\ \bibinfo
  {author} {\bibfnamefont {J.}~\bibnamefont {Faist}},\ }\href
  {https://doi.org/10.1126/science.adj3858} {\bibfield  {journal} {\bibinfo
  {journal} {Science}\ }\textbf {\bibinfo {volume} {382}},\ \bibinfo {pages}
  {434} (\bibinfo {year} {2023})}\BibitemShut {NoStop}%
\bibitem [{\citenamefont {Senanian}\ \emph {et~al.}(2023)\citenamefont
  {Senanian}, \citenamefont {Wright}, \citenamefont {Wade}, \citenamefont
  {Doyle},\ and\ \citenamefont {McMahon}}]{Senanian2023}%
  \BibitemOpen
  \bibfield  {author} {\bibinfo {author} {\bibfnamefont {A.}~\bibnamefont
  {Senanian}}, \bibinfo {author} {\bibfnamefont {L.~G.}\ \bibnamefont
  {Wright}}, \bibinfo {author} {\bibfnamefont {P.~F.}\ \bibnamefont {Wade}},
  \bibinfo {author} {\bibfnamefont {H.~K.}\ \bibnamefont {Doyle}},\ and\
  \bibinfo {author} {\bibfnamefont {P.~L.}\ \bibnamefont {McMahon}},\ }\href
  {https://doi.org/10.1038/s41567-023-02075-7} {\bibfield  {journal} {\bibinfo
  {journal} {Nature Physics}\ }\textbf {\bibinfo {volume} {19}},\ \bibinfo
  {pages} {1333} (\bibinfo {year} {2023})}\BibitemShut {NoStop}%
\bibitem [{\citenamefont {Yu}\ \emph {et~al.}(2023)\citenamefont {Yu},
  \citenamefont {Li}, \citenamefont {Wang}, \citenamefont {Leykam},
  \citenamefont {Yuan},\ and\ \citenamefont {Chen}}]{PhysRevLett.130.143801}%
  \BibitemOpen
  \bibfield  {author} {\bibinfo {author} {\bibfnamefont {D.}~\bibnamefont
  {Yu}}, \bibinfo {author} {\bibfnamefont {G.}~\bibnamefont {Li}}, \bibinfo
  {author} {\bibfnamefont {L.}~\bibnamefont {Wang}}, \bibinfo {author}
  {\bibfnamefont {D.}~\bibnamefont {Leykam}}, \bibinfo {author} {\bibfnamefont
  {L.}~\bibnamefont {Yuan}},\ and\ \bibinfo {author} {\bibfnamefont
  {X.}~\bibnamefont {Chen}},\ }\href
  {https://doi.org/10.1103/PhysRevLett.130.143801} {\bibfield  {journal}
  {\bibinfo  {journal} {Phys. Rev. Lett.}\ }\textbf {\bibinfo {volume} {130}},\
  \bibinfo {pages} {143801} (\bibinfo {year} {2023})}\BibitemShut {NoStop}%
\bibitem [{\citenamefont {Cheng}\ \emph
  {et~al.}(2023{\natexlab{b}})\citenamefont {Cheng}, \citenamefont {Wang},\
  and\ \citenamefont {Fan}}]{PhysRevLett.130.083601}%
  \BibitemOpen
  \bibfield  {author} {\bibinfo {author} {\bibfnamefont {D.}~\bibnamefont
  {Cheng}}, \bibinfo {author} {\bibfnamefont {K.}~\bibnamefont {Wang}},\ and\
  \bibinfo {author} {\bibfnamefont {S.}~\bibnamefont {Fan}},\ }\href
  {https://doi.org/10.1103/PhysRevLett.130.083601} {\bibfield  {journal}
  {\bibinfo  {journal} {Phys. Rev. Lett.}\ }\textbf {\bibinfo {volume} {130}},\
  \bibinfo {pages} {083601} (\bibinfo {year} {2023}{\natexlab{b}})}\BibitemShut
  {NoStop}%
\bibitem [{\citenamefont {Li}\ \emph {et~al.}(2023)\citenamefont {Li},
  \citenamefont {Wang}, \citenamefont {Ye}, \citenamefont {Zheng},
  \citenamefont {Wang}, \citenamefont {Liu}, \citenamefont {Dutt},
  \citenamefont {Yuan},\ and\ \citenamefont {Chen}}]{Li2023}%
  \BibitemOpen
  \bibfield  {author} {\bibinfo {author} {\bibfnamefont {G.}~\bibnamefont
  {Li}}, \bibinfo {author} {\bibfnamefont {L.}~\bibnamefont {Wang}}, \bibinfo
  {author} {\bibfnamefont {R.}~\bibnamefont {Ye}}, \bibinfo {author}
  {\bibfnamefont {Y.}~\bibnamefont {Zheng}}, \bibinfo {author} {\bibfnamefont
  {D.-W.}\ \bibnamefont {Wang}}, \bibinfo {author} {\bibfnamefont {X.-J.}\
  \bibnamefont {Liu}}, \bibinfo {author} {\bibfnamefont {A.}~\bibnamefont
  {Dutt}}, \bibinfo {author} {\bibfnamefont {L.}~\bibnamefont {Yuan}},\ and\
  \bibinfo {author} {\bibfnamefont {X.}~\bibnamefont {Chen}},\ }\href
  {https://doi.org/10.1038/s41377-023-01126-1} {\bibfield  {journal} {\bibinfo
  {journal} {Light: Science {\&} Applications}\ }\textbf {\bibinfo {volume}
  {12}},\ \bibinfo {pages} {81} (\bibinfo {year} {2023})}\BibitemShut {NoStop}%
\bibitem [{\citenamefont {Dutt}\ \emph {et~al.}(2022)\citenamefont {Dutt},
  \citenamefont {Yuan}, \citenamefont {Yang}, \citenamefont {Wang},
  \citenamefont {Buddhiraju}, \citenamefont {Vučković},\ and\ \citenamefont
  {Fan}}]{dutt_creating_2022}%
  \BibitemOpen
  \bibfield  {author} {\bibinfo {author} {\bibfnamefont {A.}~\bibnamefont
  {Dutt}}, \bibinfo {author} {\bibfnamefont {L.}~\bibnamefont {Yuan}}, \bibinfo
  {author} {\bibfnamefont {K.~Y.}\ \bibnamefont {Yang}}, \bibinfo {author}
  {\bibfnamefont {K.}~\bibnamefont {Wang}}, \bibinfo {author} {\bibfnamefont
  {S.}~\bibnamefont {Buddhiraju}}, \bibinfo {author} {\bibfnamefont
  {J.}~\bibnamefont {Vučković}},\ and\ \bibinfo {author} {\bibfnamefont
  {S.}~\bibnamefont {Fan}},\ }\href
  {https://doi.org/10.1038/s41467-022-31140-7} {\bibfield  {journal} {\bibinfo
  {journal} {Nature Communications}\ }\textbf {\bibinfo {volume} {13}},\
  \bibinfo {pages} {3377} (\bibinfo {year} {2022})},\ \bibinfo {note} {number:
  1 Publisher: Nature Publishing Group}\BibitemShut {NoStop}%
\bibitem [{\citenamefont {Nation}\ and\ \citenamefont
  {Porras}(2018)}]{Nation_2018}%
  \BibitemOpen
  \bibfield  {author} {\bibinfo {author} {\bibfnamefont {C.}~\bibnamefont
  {Nation}}\ and\ \bibinfo {author} {\bibfnamefont {D.}~\bibnamefont
  {Porras}},\ }\href {https://doi.org/10.1088/1367-2630/aae28f} {\bibfield
  {journal} {\bibinfo  {journal} {New Journal of Physics}\ }\textbf {\bibinfo
  {volume} {20}},\ \bibinfo {pages} {103003} (\bibinfo {year}
  {2018})}\BibitemShut {NoStop}%
\bibitem [{\citenamefont {G{\'o}mez}\ \emph {et~al.}(2011)\citenamefont
  {G{\'o}mez}, \citenamefont {Kar}, \citenamefont {Kota}, \citenamefont
  {Molina}, \citenamefont {Rela{\~{n}}o},\ and\ \citenamefont
  {Retamosa}}]{gomez2011many}%
  \BibitemOpen
  \bibfield  {author} {\bibinfo {author} {\bibfnamefont {J.~M.~G.}\
  \bibnamefont {G{\'o}mez}}, \bibinfo {author} {\bibfnamefont {K.}~\bibnamefont
  {Kar}}, \bibinfo {author} {\bibfnamefont {V.~K.~B.}\ \bibnamefont {Kota}},
  \bibinfo {author} {\bibfnamefont {R.~A.}\ \bibnamefont {Molina}}, \bibinfo
  {author} {\bibfnamefont {A.}~\bibnamefont {Rela{\~{n}}o}},\ and\ \bibinfo
  {author} {\bibfnamefont {J.}~\bibnamefont {Retamosa}},\ }\href
  {https://www.sciencedirect.com/science/article/pii/S0370157310003091}
  {\bibfield  {journal} {\bibinfo  {journal} {Physics Reports}\ }\textbf
  {\bibinfo {volume} {499}},\ \bibinfo {pages} {103} (\bibinfo {year}
  {2011})}\BibitemShut {NoStop}%
\bibitem [{\citenamefont {Kos}\ \emph {et~al.}(2018)\citenamefont {Kos},
  \citenamefont {Ljubotina},\ and\ \citenamefont {Prosen}}]{PhysRevX.8.021062}%
  \BibitemOpen
  \bibfield  {author} {\bibinfo {author} {\bibfnamefont {P.}~\bibnamefont
  {Kos}}, \bibinfo {author} {\bibfnamefont {M.}~\bibnamefont {Ljubotina}},\
  and\ \bibinfo {author} {\bibfnamefont {T.~c.~v.}\ \bibnamefont {Prosen}},\
  }\href {https://doi.org/10.1103/PhysRevX.8.021062} {\bibfield  {journal}
  {\bibinfo  {journal} {Phys. Rev. X}\ }\textbf {\bibinfo {volume} {8}},\
  \bibinfo {pages} {021062} (\bibinfo {year} {2018})}\BibitemShut {NoStop}%
\bibitem [{\citenamefont {St{\"o}ckmann}(2000)}]{stockmann2000quantum}%
  \BibitemOpen
  \bibfield  {author} {\bibinfo {author} {\bibfnamefont {H.-J.}\ \bibnamefont
  {St{\"o}ckmann}},\ }\href {https://doi.org/10.1017/CBO9780511524622}
  {\bibinfo {title} {Quantum chaos: an introduction}} (\bibinfo {year}
  {2000})\BibitemShut {NoStop}%
\bibitem [{\citenamefont {Pal}\ and\ \citenamefont {Huse}(2010)}]{pal2010many}%
  \BibitemOpen
  \bibfield  {author} {\bibinfo {author} {\bibfnamefont {A.}~\bibnamefont
  {Pal}}\ and\ \bibinfo {author} {\bibfnamefont {D.~A.}\ \bibnamefont {Huse}},\
  }\href {https://doi.org/10.1103/PhysRevB.82.174411} {\bibfield  {journal}
  {\bibinfo  {journal} {Phys. Rev. B}\ }\textbf {\bibinfo {volume} {82}},\
  \bibinfo {pages} {174411} (\bibinfo {year} {2010})}\BibitemShut {NoStop}%
\bibitem [{\citenamefont {Nandkishore}\ and\ \citenamefont
  {Huse}(2015)}]{nandkishore2015many}%
  \BibitemOpen
  \bibfield  {author} {\bibinfo {author} {\bibfnamefont {R.}~\bibnamefont
  {Nandkishore}}\ and\ \bibinfo {author} {\bibfnamefont {D.~A.}\ \bibnamefont
  {Huse}},\ }\href {https://doi.org/10.1146/annurev-conmatphys-031214-014726}
  {\bibfield  {journal} {\bibinfo  {journal} {Annu. Rev. Condens. Matter
  Phys.}\ }\textbf {\bibinfo {volume} {6}},\ \bibinfo {pages} {15} (\bibinfo
  {year} {2015})}\BibitemShut {NoStop}%
\bibitem [{\citenamefont {Smith}\ \emph {et~al.}(2016)\citenamefont {Smith},
  \citenamefont {Lee}, \citenamefont {Richerme}, \citenamefont {Neyenhuis},
  \citenamefont {Hess}, \citenamefont {Hauke}, \citenamefont {Heyl},
  \citenamefont {Huse},\ and\ \citenamefont {Monroe}}]{smith2016many}%
  \BibitemOpen
  \bibfield  {author} {\bibinfo {author} {\bibfnamefont {J.}~\bibnamefont
  {Smith}}, \bibinfo {author} {\bibfnamefont {A.}~\bibnamefont {Lee}}, \bibinfo
  {author} {\bibfnamefont {P.}~\bibnamefont {Richerme}}, \bibinfo {author}
  {\bibfnamefont {B.}~\bibnamefont {Neyenhuis}}, \bibinfo {author}
  {\bibfnamefont {P.~W.}\ \bibnamefont {Hess}}, \bibinfo {author}
  {\bibfnamefont {P.}~\bibnamefont {Hauke}}, \bibinfo {author} {\bibfnamefont
  {M.}~\bibnamefont {Heyl}}, \bibinfo {author} {\bibfnamefont {D.~A.}\
  \bibnamefont {Huse}},\ and\ \bibinfo {author} {\bibfnamefont
  {C.}~\bibnamefont {Monroe}},\ }\href {https://doi.org/10.1038/nphys3783}
  {\bibfield  {journal} {\bibinfo  {journal} {Nature Physics}\ }\textbf
  {\bibinfo {volume} {12}},\ \bibinfo {pages} {907} (\bibinfo {year}
  {2016})}\BibitemShut {NoStop}%
\bibitem [{\citenamefont {Abanin}\ \emph {et~al.}(2019)\citenamefont {Abanin},
  \citenamefont {Altman}, \citenamefont {Bloch},\ and\ \citenamefont
  {Serbyn}}]{abanin2019colloquium}%
  \BibitemOpen
  \bibfield  {author} {\bibinfo {author} {\bibfnamefont {D.~A.}\ \bibnamefont
  {Abanin}}, \bibinfo {author} {\bibfnamefont {E.}~\bibnamefont {Altman}},
  \bibinfo {author} {\bibfnamefont {I.}~\bibnamefont {Bloch}},\ and\ \bibinfo
  {author} {\bibfnamefont {M.}~\bibnamefont {Serbyn}},\ }\href
  {https://doi.org/10.1103/RevModPhys.91.021001} {\bibfield  {journal}
  {\bibinfo  {journal} {Rev. Mod. Phys.}\ }\textbf {\bibinfo {volume} {91}},\
  \bibinfo {pages} {021001} (\bibinfo {year} {2019})}\BibitemShut {NoStop}%
\bibitem [{\citenamefont {Alet}\ and\ \citenamefont
  {Laflorencie}(2018)}]{alet2018many}%
  \BibitemOpen
  \bibfield  {author} {\bibinfo {author} {\bibfnamefont {F.}~\bibnamefont
  {Alet}}\ and\ \bibinfo {author} {\bibfnamefont {N.}~\bibnamefont
  {Laflorencie}},\ }\href
  {https://www.sciencedirect.com/science/article/pii/S163107051830032X}
  {\bibfield  {journal} {\bibinfo  {journal} {Comptes Rendus Physique}\
  }\textbf {\bibinfo {volume} {19}},\ \bibinfo {pages} {498} (\bibinfo {year}
  {2018})}\BibitemShut {NoStop}%
\bibitem [{\citenamefont {Roy}\ \emph {et~al.}(2022)\citenamefont {Roy},
  \citenamefont {Mishra},\ and\ \citenamefont {Prosen}}]{PhysRevE.106.024208}%
  \BibitemOpen
  \bibfield  {author} {\bibinfo {author} {\bibfnamefont {D.}~\bibnamefont
  {Roy}}, \bibinfo {author} {\bibfnamefont {D.}~\bibnamefont {Mishra}},\ and\
  \bibinfo {author} {\bibfnamefont {T.~c.~v.}\ \bibnamefont {Prosen}},\ }\href
  {https://doi.org/10.1103/PhysRevE.106.024208} {\bibfield  {journal} {\bibinfo
   {journal} {Phys. Rev. E}\ }\textbf {\bibinfo {volume} {106}},\ \bibinfo
  {pages} {024208} (\bibinfo {year} {2022})}\BibitemShut {NoStop}%
\bibitem [{\citenamefont {Joshi}\ \emph {et~al.}(2022)\citenamefont {Joshi},
  \citenamefont {Elben}, \citenamefont {Vikram}, \citenamefont {Vermersch},
  \citenamefont {Galitski},\ and\ \citenamefont {Zoller}}]{joshi_probing_2022}%
  \BibitemOpen
  \bibfield  {author} {\bibinfo {author} {\bibfnamefont {L.~K.}\ \bibnamefont
  {Joshi}}, \bibinfo {author} {\bibfnamefont {A.}~\bibnamefont {Elben}},
  \bibinfo {author} {\bibfnamefont {A.}~\bibnamefont {Vikram}}, \bibinfo
  {author} {\bibfnamefont {B.}~\bibnamefont {Vermersch}}, \bibinfo {author}
  {\bibfnamefont {V.}~\bibnamefont {Galitski}},\ and\ \bibinfo {author}
  {\bibfnamefont {P.}~\bibnamefont {Zoller}},\ }\href
  {https://doi.org/10.1103/PhysRevX.12.011018} {\bibfield  {journal} {\bibinfo
  {journal} {Physical Review X}\ }\textbf {\bibinfo {volume} {12}},\ \bibinfo
  {pages} {011018} (\bibinfo {year} {2022})},\ \bibinfo {note} {publisher:
  American Physical Society}\BibitemShut {NoStop}%
\bibitem [{\citenamefont {Yuan}\ \emph {et~al.}(2019)\citenamefont {Yuan},
  \citenamefont {Lin}, \citenamefont {Zhang}, \citenamefont {Xiao},
  \citenamefont {Chen},\ and\ \citenamefont {Fan}}]{PhysRevLett.122.083903}%
  \BibitemOpen
  \bibfield  {author} {\bibinfo {author} {\bibfnamefont {L.}~\bibnamefont
  {Yuan}}, \bibinfo {author} {\bibfnamefont {Q.}~\bibnamefont {Lin}}, \bibinfo
  {author} {\bibfnamefont {A.}~\bibnamefont {Zhang}}, \bibinfo {author}
  {\bibfnamefont {M.}~\bibnamefont {Xiao}}, \bibinfo {author} {\bibfnamefont
  {X.}~\bibnamefont {Chen}},\ and\ \bibinfo {author} {\bibfnamefont
  {S.}~\bibnamefont {Fan}},\ }\href
  {https://doi.org/10.1103/PhysRevLett.122.083903} {\bibfield  {journal}
  {\bibinfo  {journal} {Phys. Rev. Lett.}\ }\textbf {\bibinfo {volume} {122}},\
  \bibinfo {pages} {083903} (\bibinfo {year} {2019})}\BibitemShut {NoStop}%
\bibitem [{\citenamefont {Yariv}(2007)}]{Optical_Electronics_in_Modern}%
  \BibitemOpen
  \bibfield  {author} {\bibinfo {author} {\bibfnamefont {P.~Y.}\ \bibnamefont
  {Yariv}, \bibfnamefont {Amnon}},\ }\href@noop {} {\emph {\bibinfo {title}
  {Photonics: Optical Electronics in Modern Communications}}}\ (\bibinfo
  {publisher} {Oxford University Press},\ \bibinfo {year} {2007})\ \bibinfo
  {note} {publisher: Oxford University Press}\BibitemShut {NoStop}%
\bibitem [{\citenamefont {Strekalov}\ \emph {et~al.}(2016)\citenamefont
  {Strekalov}, \citenamefont {Marquardt}, \citenamefont {Matsko}, \citenamefont
  {Schwefel},\ and\ \citenamefont {Leuchs}}]{strekalov_nonlinear_2016}%
  \BibitemOpen
  \bibfield  {author} {\bibinfo {author} {\bibfnamefont {D.~V.}\ \bibnamefont
  {Strekalov}}, \bibinfo {author} {\bibfnamefont {C.}~\bibnamefont
  {Marquardt}}, \bibinfo {author} {\bibfnamefont {A.~B.}\ \bibnamefont
  {Matsko}}, \bibinfo {author} {\bibfnamefont {H.~G.~L.}\ \bibnamefont
  {Schwefel}},\ and\ \bibinfo {author} {\bibfnamefont {G.}~\bibnamefont
  {Leuchs}},\ }\href {https://doi.org/10.1088/2040-8978/18/12/123002}
  {\bibfield  {journal} {\bibinfo  {journal} {Journal of Optics}\ }\textbf
  {\bibinfo {volume} {18}},\ \bibinfo {pages} {123002} (\bibinfo {year}
  {2016})},\ \bibinfo {note} {publisher: IOP Publishing}\BibitemShut {NoStop}%
\bibitem [{\citenamefont {Kriecherbauer}\ \emph {et~al.}(2001)\citenamefont
  {Kriecherbauer}, \citenamefont {Marklof},\ and\ \citenamefont
  {Soshnikov}}]{kriecherbauer_random_2001}%
  \BibitemOpen
  \bibfield  {author} {\bibinfo {author} {\bibfnamefont {T.}~\bibnamefont
  {Kriecherbauer}}, \bibinfo {author} {\bibfnamefont {J.}~\bibnamefont
  {Marklof}},\ and\ \bibinfo {author} {\bibfnamefont {A.}~\bibnamefont
  {Soshnikov}},\ }\href {https://doi.org/10.1073/pnas.191366198} {\bibfield
  {journal} {\bibinfo  {journal} {Proceedings of the National Academy of
  Sciences}\ }\textbf {\bibinfo {volume} {98}},\ \bibinfo {pages} {10531}
  (\bibinfo {year} {2001})},\ \bibinfo {note} {publisher: Proceedings of the
  National Academy of Sciences}\BibitemShut {NoStop}%
\bibitem [{\citenamefont {Oganesyan}\ and\ \citenamefont
  {Huse}(2007)}]{PhysRevB.75.155111}%
  \BibitemOpen
  \bibfield  {author} {\bibinfo {author} {\bibfnamefont {V.}~\bibnamefont
  {Oganesyan}}\ and\ \bibinfo {author} {\bibfnamefont {D.~A.}\ \bibnamefont
  {Huse}},\ }\href {https://doi.org/10.1103/PhysRevB.75.155111} {\bibfield
  {journal} {\bibinfo  {journal} {Phys. Rev. B}\ }\textbf {\bibinfo {volume}
  {75}},\ \bibinfo {pages} {155111} (\bibinfo {year} {2007})}\BibitemShut
  {NoStop}%
\bibitem [{\citenamefont {Wannier}(1960)}]{wannier_wave_1960}%
  \BibitemOpen
  \bibfield  {author} {\bibinfo {author} {\bibfnamefont {G.~H.}\ \bibnamefont
  {Wannier}},\ }\href {https://doi.org/10.1103/PhysRev.117.432} {\bibfield
  {journal} {\bibinfo  {journal} {Physical Review}\ }\textbf {\bibinfo {volume}
  {117}},\ \bibinfo {pages} {432} (\bibinfo {year} {1960})},\ \bibinfo {note}
  {publisher: American Physical Society}\BibitemShut {NoStop}%
\bibitem [{\citenamefont {van Nieuwenburg}\ \emph {et~al.}(2019)\citenamefont
  {van Nieuwenburg}, \citenamefont {Baum},\ and\ \citenamefont
  {Refael}}]{van_nieuwenburg_bloch_2019}%
  \BibitemOpen
  \bibfield  {author} {\bibinfo {author} {\bibfnamefont {E.}~\bibnamefont {van
  Nieuwenburg}}, \bibinfo {author} {\bibfnamefont {Y.}~\bibnamefont {Baum}},\
  and\ \bibinfo {author} {\bibfnamefont {G.}~\bibnamefont {Refael}},\ }\href
  {https://doi.org/10.1073/pnas.1819316116} {\bibfield  {journal} {\bibinfo
  {journal} {Proceedings of the National Academy of Sciences}\ }\textbf
  {\bibinfo {volume} {116}},\ \bibinfo {pages} {9269} (\bibinfo {year}
  {2019})},\ \bibinfo {note} {publisher: Proceedings of the National Academy of
  Sciences}\BibitemShut {NoStop}%
\bibitem [{\citenamefont {Amico}\ \emph {et~al.}(2008)\citenamefont {Amico},
  \citenamefont {Fazio}, \citenamefont {Osterloh},\ and\ \citenamefont
  {Vedral}}]{amico_entanglement_2008}%
  \BibitemOpen
  \bibfield  {author} {\bibinfo {author} {\bibfnamefont {L.}~\bibnamefont
  {Amico}}, \bibinfo {author} {\bibfnamefont {R.}~\bibnamefont {Fazio}},
  \bibinfo {author} {\bibfnamefont {A.}~\bibnamefont {Osterloh}},\ and\
  \bibinfo {author} {\bibfnamefont {V.}~\bibnamefont {Vedral}},\ }\href
  {https://doi.org/10.1103/RevModPhys.80.517} {\bibfield  {journal} {\bibinfo
  {journal} {Reviews of Modern Physics}\ }\textbf {\bibinfo {volume} {80}},\
  \bibinfo {pages} {517} (\bibinfo {year} {2008})},\ \bibinfo {note}
  {publisher: American Physical Society}\BibitemShut {NoStop}%
\bibitem [{\citenamefont {Li}\ \emph {et~al.}(2001)\citenamefont {Li},
  \citenamefont {Zeng}, \citenamefont {Liu},\ and\ \citenamefont
  {Long}}]{li_entanglement_2001}%
  \BibitemOpen
  \bibfield  {author} {\bibinfo {author} {\bibfnamefont {Y.~S.}\ \bibnamefont
  {Li}}, \bibinfo {author} {\bibfnamefont {B.}~\bibnamefont {Zeng}}, \bibinfo
  {author} {\bibfnamefont {X.~S.}\ \bibnamefont {Liu}},\ and\ \bibinfo {author}
  {\bibfnamefont {G.~L.}\ \bibnamefont {Long}},\ }\href
  {https://doi.org/10.1103/PhysRevA.64.054302} {\bibfield  {journal} {\bibinfo
  {journal} {Physical Review A}\ }\textbf {\bibinfo {volume} {64}},\ \bibinfo
  {pages} {054302} (\bibinfo {year} {2001})},\ \bibinfo {note} {publisher:
  American Physical Society}\BibitemShut {NoStop}%
\bibitem [{\citenamefont {Schulz}\ \emph {et~al.}(2019)\citenamefont {Schulz},
  \citenamefont {Hooley}, \citenamefont {Moessner},\ and\ \citenamefont
  {Pollmann}}]{schulz_stark_2019}%
  \BibitemOpen
  \bibfield  {author} {\bibinfo {author} {\bibfnamefont {M.}~\bibnamefont
  {Schulz}}, \bibinfo {author} {\bibfnamefont {C.}~\bibnamefont {Hooley}},
  \bibinfo {author} {\bibfnamefont {R.}~\bibnamefont {Moessner}},\ and\
  \bibinfo {author} {\bibfnamefont {F.}~\bibnamefont {Pollmann}},\ }\href
  {https://doi.org/10.1103/PhysRevLett.122.040606} {\bibfield  {journal}
  {\bibinfo  {journal} {Physical Review Letters}\ }\textbf {\bibinfo {volume}
  {122}},\ \bibinfo {pages} {040606} (\bibinfo {year} {2019})},\ \bibinfo
  {note} {publisher: American Physical Society}\BibitemShut {NoStop}%
\bibitem [{\citenamefont {Kaneda}\ \emph {et~al.}(2019)\citenamefont {Kaneda},
  \citenamefont {Suzuki}, \citenamefont {Shimizu},\ and\ \citenamefont
  {Edamatsu}}]{frequency-bin}%
  \BibitemOpen
  \bibfield  {author} {\bibinfo {author} {\bibfnamefont {F.}~\bibnamefont
  {Kaneda}}, \bibinfo {author} {\bibfnamefont {H.}~\bibnamefont {Suzuki}},
  \bibinfo {author} {\bibfnamefont {R.}~\bibnamefont {Shimizu}},\ and\ \bibinfo
  {author} {\bibfnamefont {K.}~\bibnamefont {Edamatsu}},\ }\href
  {https://doi.org/10.1364/OE.27.001416} {\bibfield  {journal} {\bibinfo
  {journal} {Opt. Express}\ }\textbf {\bibinfo {volume} {27}},\ \bibinfo
  {pages} {1416} (\bibinfo {year} {2019})}\BibitemShut {NoStop}%
\bibitem [{\citenamefont {Olislager}\ \emph {et~al.}(2010)\citenamefont
  {Olislager}, \citenamefont {Cussey}, \citenamefont {Nguyen}, \citenamefont
  {Emplit}, \citenamefont {Massar}, \citenamefont {Merolla},\ and\
  \citenamefont {Huy}}]{PRA-frequency-bin}%
  \BibitemOpen
  \bibfield  {author} {\bibinfo {author} {\bibfnamefont {L.}~\bibnamefont
  {Olislager}}, \bibinfo {author} {\bibfnamefont {J.}~\bibnamefont {Cussey}},
  \bibinfo {author} {\bibfnamefont {A.~T.}\ \bibnamefont {Nguyen}}, \bibinfo
  {author} {\bibfnamefont {P.}~\bibnamefont {Emplit}}, \bibinfo {author}
  {\bibfnamefont {S.}~\bibnamefont {Massar}}, \bibinfo {author} {\bibfnamefont
  {J.-M.}\ \bibnamefont {Merolla}},\ and\ \bibinfo {author} {\bibfnamefont
  {K.~P.}\ \bibnamefont {Huy}},\ }\href
  {https://doi.org/10.1103/PhysRevA.82.013804} {\bibfield  {journal} {\bibinfo
  {journal} {Phys. Rev. A}\ }\textbf {\bibinfo {volume} {82}},\ \bibinfo
  {pages} {013804} (\bibinfo {year} {2010})}\BibitemShut {NoStop}%
\bibitem [{Rid(2023)}]{Rid}%
  \BibitemOpen
  \href {https://refractiveindex.info/} {\bibinfo {title} {Refractive index
  database}} (\bibinfo {year} {2023})\BibitemShut {NoStop}%
\bibitem [{\citenamefont {Dutt}\ \emph {et~al.}(2020)\citenamefont {Dutt},
  \citenamefont {Minkov}, \citenamefont {Williamson},\ and\ \citenamefont
  {Fan}}]{Dutt2020}%
  \BibitemOpen
  \bibfield  {author} {\bibinfo {author} {\bibfnamefont {A.}~\bibnamefont
  {Dutt}}, \bibinfo {author} {\bibfnamefont {M.}~\bibnamefont {Minkov}},
  \bibinfo {author} {\bibfnamefont {I.~A.~D.}\ \bibnamefont {Williamson}},\
  and\ \bibinfo {author} {\bibfnamefont {S.}~\bibnamefont {Fan}},\ }\href
  {https://doi.org/10.1038/s41377-020-0334-8} {\bibfield  {journal} {\bibinfo
  {journal} {Light: Science {\&} Applications}\ }\textbf {\bibinfo {volume}
  {9}},\ \bibinfo {pages} {131} (\bibinfo {year} {2020})}\BibitemShut {NoStop}%
\bibitem [{\citenamefont {Juki\ifmmode~\acute{c}\else \'{c}\fi{}}\ and\
  \citenamefont {Buljan}(2013)}]{jukic2013four}%
  \BibitemOpen
  \bibfield  {author} {\bibinfo {author} {\bibfnamefont {D.}~\bibnamefont
  {Juki\ifmmode~\acute{c}\else \'{c}\fi{}}}\ and\ \bibinfo {author}
  {\bibfnamefont {H.}~\bibnamefont {Buljan}},\ }\href
  {https://doi.org/10.1103/PhysRevA.87.013814} {\bibfield  {journal} {\bibinfo
  {journal} {Phys. Rev. A}\ }\textbf {\bibinfo {volume} {87}},\ \bibinfo
  {pages} {013814} (\bibinfo {year} {2013})}\BibitemShut {NoStop}%
\bibitem [{\citenamefont {Lustig}\ \emph
  {et~al.}(2019{\natexlab{b}})\citenamefont {Lustig}, \citenamefont {Weimann},
  \citenamefont {Plotnik}, \citenamefont {Lumer}, \citenamefont {Bandres},
  \citenamefont {Szameit},\ and\ \citenamefont {Segev}}]{lustig2019photonic}%
  \BibitemOpen
  \bibfield  {author} {\bibinfo {author} {\bibfnamefont {E.}~\bibnamefont
  {Lustig}}, \bibinfo {author} {\bibfnamefont {S.}~\bibnamefont {Weimann}},
  \bibinfo {author} {\bibfnamefont {Y.}~\bibnamefont {Plotnik}}, \bibinfo
  {author} {\bibfnamefont {Y.}~\bibnamefont {Lumer}}, \bibinfo {author}
  {\bibfnamefont {M.~A.}\ \bibnamefont {Bandres}}, \bibinfo {author}
  {\bibfnamefont {A.}~\bibnamefont {Szameit}},\ and\ \bibinfo {author}
  {\bibfnamefont {M.}~\bibnamefont {Segev}},\ }\href
  {https://doi.org/10.1038/s41586-019-0943-7} {\bibfield  {journal} {\bibinfo
  {journal} {Nature}\ }\textbf {\bibinfo {volume} {567}},\ \bibinfo {pages}
  {356} (\bibinfo {year} {2019}{\natexlab{b}})}\BibitemShut {NoStop}%
\bibitem [{\citenamefont {Boada}\ \emph {et~al.}(2012)\citenamefont {Boada},
  \citenamefont {Celi}, \citenamefont {Latorre},\ and\ \citenamefont
  {Lewenstein}}]{boada2012quantum}%
  \BibitemOpen
  \bibfield  {author} {\bibinfo {author} {\bibfnamefont {O.}~\bibnamefont
  {Boada}}, \bibinfo {author} {\bibfnamefont {A.}~\bibnamefont {Celi}},
  \bibinfo {author} {\bibfnamefont {J.~I.}\ \bibnamefont {Latorre}},\ and\
  \bibinfo {author} {\bibfnamefont {M.}~\bibnamefont {Lewenstein}},\ }\href
  {https://doi.org/10.1103/PhysRevLett.108.133001} {\bibfield  {journal}
  {\bibinfo  {journal} {Phys. Rev. Lett.}\ }\textbf {\bibinfo {volume} {108}},\
  \bibinfo {pages} {133001} (\bibinfo {year} {2012})}\BibitemShut {NoStop}%
\bibitem [{\citenamefont {Scherg}\ \emph {et~al.}(2021)\citenamefont {Scherg},
  \citenamefont {Kohlert}, \citenamefont {Sala}, \citenamefont {Pollmann},
  \citenamefont {Hebbe~Madhusudhana}, \citenamefont {Bloch},\ and\
  \citenamefont {Aidelsburger}}]{scherg_observing_2021}%
  \BibitemOpen
  \bibfield  {author} {\bibinfo {author} {\bibfnamefont {S.}~\bibnamefont
  {Scherg}}, \bibinfo {author} {\bibfnamefont {T.}~\bibnamefont {Kohlert}},
  \bibinfo {author} {\bibfnamefont {P.}~\bibnamefont {Sala}}, \bibinfo {author}
  {\bibfnamefont {F.}~\bibnamefont {Pollmann}}, \bibinfo {author}
  {\bibfnamefont {B.}~\bibnamefont {Hebbe~Madhusudhana}}, \bibinfo {author}
  {\bibfnamefont {I.}~\bibnamefont {Bloch}},\ and\ \bibinfo {author}
  {\bibfnamefont {M.}~\bibnamefont {Aidelsburger}},\ }\href
  {https://doi.org/10.1038/s41467-021-24726-0} {\bibfield  {journal} {\bibinfo
  {journal} {Nature Communications}\ }\textbf {\bibinfo {volume} {12}},\
  \bibinfo {pages} {4490} (\bibinfo {year} {2021})},\ \bibinfo {note} {number:
  1 Publisher: Nature Publishing Group}\BibitemShut {NoStop}%
\bibitem [{\citenamefont {Yang}\ \emph {et~al.}(2018)\citenamefont {Yang},
  \citenamefont {Tang},\ and\ \citenamefont {Jacob}}]{PhysRevA.Markovian}%
  \BibitemOpen
  \bibfield  {author} {\bibinfo {author} {\bibfnamefont {L.-P.}\ \bibnamefont
  {Yang}}, \bibinfo {author} {\bibfnamefont {H.~X.}\ \bibnamefont {Tang}},\
  and\ \bibinfo {author} {\bibfnamefont {Z.}~\bibnamefont {Jacob}},\ }\href
  {https://doi.org/10.1103/PhysRevA.97.013833} {\bibfield  {journal} {\bibinfo
  {journal} {Phys. Rev. A}\ }\textbf {\bibinfo {volume} {97}},\ \bibinfo
  {pages} {013833} (\bibinfo {year} {2018})}\BibitemShut {NoStop}%
\bibitem [{\citenamefont {Breuer}\ and\ \citenamefont
  {Petruccione}(2007)}]{Breuer2007}%
  \BibitemOpen
  \bibfield  {author} {\bibinfo {author} {\bibfnamefont {H.-P.}\ \bibnamefont
  {Breuer}}\ and\ \bibinfo {author} {\bibfnamefont {F.}~\bibnamefont
  {Petruccione}},\ }\href
  {https://doi.org/10.1093/acprof:oso/9780199213900.001.0001} {\bibinfo {title}
  {The theory of open quantum systems}} (\bibinfo {year} {2007})\BibitemShut
  {NoStop}%
\bibitem [{\citenamefont {Shen}\ and\ \citenamefont
  {Fan}(2009)}]{PhysRevA.79.023838}%
  \BibitemOpen
  \bibfield  {author} {\bibinfo {author} {\bibfnamefont {J.-T.}\ \bibnamefont
  {Shen}}\ and\ \bibinfo {author} {\bibfnamefont {S.}~\bibnamefont {Fan}},\
  }\href {https://doi.org/10.1103/PhysRevA.79.023838} {\bibfield  {journal}
  {\bibinfo  {journal} {Phys. Rev. A}\ }\textbf {\bibinfo {volume} {79}},\
  \bibinfo {pages} {023838} (\bibinfo {year} {2009})}\BibitemShut {NoStop}%
\bibitem [{\citenamefont {Fan}\ \emph {et~al.}(2010)\citenamefont {Fan},
  \citenamefont {Kocaba\ifmmode~\mbox{\c{s}}\else \c{s}\fi{}},\ and\
  \citenamefont {Shen}}]{PhysRevA.82.063821}%
  \BibitemOpen
  \bibfield  {author} {\bibinfo {author} {\bibfnamefont {S.}~\bibnamefont
  {Fan}}, \bibinfo {author} {\bibfnamefont {i.~m. c.~E.}\ \bibnamefont
  {Kocaba\ifmmode~\mbox{\c{s}}\else \c{s}\fi{}}},\ and\ \bibinfo {author}
  {\bibfnamefont {J.-T.}\ \bibnamefont {Shen}},\ }\href
  {https://doi.org/10.1103/PhysRevA.82.063821} {\bibfield  {journal} {\bibinfo
  {journal} {Phys. Rev. A}\ }\textbf {\bibinfo {volume} {82}},\ \bibinfo
  {pages} {063821} (\bibinfo {year} {2010})}\BibitemShut {NoStop}%
\end{thebibliography}
\end{document}